\documentstyle[10pt,aaspp4,amssym]{article}
\input epsf
\def\Deg{${}^{\circ}$\llap{.}}
\def\Sec{${}^{\prime\prime}$\llap{.}}
\def\Min{${}^{\prime}$\llap{.}}
\lefthead{SANDQUIST, BOLTE AND STETSON.}
\righthead{CCD PHOTOMETRY IN M5}
\begin{document}
\title{CCD Photometry of the Globular Cluster M5. I. The
Color-Magnitude Diagram and Luminosity Functions}

\author{Eric L. Sandquist, Michael Bolte}
\affil{UCO/Lick Observatory, University of California, Santa Cruz,
CA~95064; erics@ucolick.org, bolte@ucolick.org}
\author{and Peter B. Stetson}
\affil{Dominion Astrophysical Observatory, Herzberg Institute of
Astrophysics, National Research Council of Canada, 5071 West Saanich Road,
Victoria, BC V8X 4M6, Canada; stetson@dao.nrc.ca}

\dates

\begin{abstract}
We present new $BVI$ photometry for the halo globular cluster M5 (NGC
5904 = C1516+022), and examine the $B$- and $I$-band luminosity
functions (LFs), based on over 20,000 stars --- one of the largest
samples ever gathered for a cluster luminosity function. Extensive
artificial star tests have been conducted to quantify incompleteness
as a function of magnitude and cluster radius. We do not see evidence
in the LF of a ``subgiant excess'' or of a discrepancy in the relative
numbers of stars on the red-giant branch and main sequence, both of
which have been claimed in more metal-poor clusters.

Enhancements of $\alpha$-element have been taken into account in our
analysis.  This improves the agreement between the observed and
predicted positions of the ``red-giant bump''.  Depending on the
average $\alpha$-element enhancement among globular clusters and the
distance calibration, the observed discrepancy between the theoretical
and observed position for a large number of clusters (Fusi Pecci et
al. 1990) can be almost completely removed.

The helium abundance of M5, as determined by the population ratio $R$,
is found to be $Y = 0.19 \pm 0.02$. However, there is no other indication
that the helium abundance is different from other clusters of
similar metallicity, and values calculated for other helium indicators
are consistent with $Y \approx 0.23$.

The relative ages of M5, Palomar 5, M4, NGC~288, NGC~362, NGC~1261,
NGC~1851 and NGC~2808 are derived via the $\Delta V_{TO}^{HB}$ method
using M5's horizontal branch (HB) as a bridge to compare clusters with
very different HB morphology.  We conclude that at the level of $\sim
1.5$ Gyr these clusters of comparable metallicity are the same age with
the possible exception of NGC~288 (older by $3.5\pm 1.5$ if the reddest
NGC~288 HB stars are on the zero-age horizontal branch) and Palomar~5
(which may be marginally younger). Even with NGC~288 set aside, there is
a large range in HB morphology between the remaining clusters which
appears to eliminate age as the sole second parameter determining HB
morphology in the case of constant mass loss between RGB and HB
(although a Reimers' mass-loss relation weakens this statement
considerably).

We are unable to chose between the two competing values for M5's
(absolute) metallicity: [Fe/H] $= -1.40$ (Zinn \& West 1984) and
$-1.17$ (Sneden et al. 1992) based on recent high-dispersion
spectroscopy. This level of discrepancy has a signifcant effect on the
derivation of the distance modulus and absolute age of M5.
From subdwarf fitting to the main sequence of the cluster, we find an apparent
distance modulus $(m - M)_{V} = 14.41 \pm 0.07$ for [Fe/H]$_{M5} =
-1.40$, and $14.50 \pm 0.07$ if [Fe/H]$_{M5} = -1.17$.
From comparisons with theoretical isochrones and
luminosity functions, we find an absolute age for M5 of $13.5 \pm 1$
Gyr (internal error, assuming perfect models and no [$M$/H] error) for the
Zinn \& West abundance scale and $11 \pm 1$ Gyr for the higher
abundance value.

\end{abstract}


\section{Introduction}\label{intro}
Stellar population studies of Galactic globular clusters (GGC) have
played a critical role in the last 40 years of efforts to understand 
stellar evolution for $\sim 1M_\odot$ stars in the metal-poor
regime. The luminosity function (LF) for evolved stars gives a
direct measure of the time stars spend in each phase of post-main-sequence
evolution, so that the LF is an expression of physical processes
occuring deep in the
interior of the stars. Renzini \& Fusi Pecci (1988) give an
excellent comprehensive review of the critical role accurate LFs
for GGC have to play in evaluating the adequacy of the canonical
evolution models and in deriving values for
key quantities such as helium abundance, the extent of convective
overshooting, neutrino cooling rates, and mass loss on the giant branches.

However, the challenges to observers presented in Renzini \& Fusi Pecci
have not yet been met. Until very recently, the only means of
measuring large samples of evolved stars in GGC was via photographic
plates. The difficulties
of accurate photometry in crowded fields
with a non-linear detector are sufficiently large that LFs including
both evolved stars and main sequence stars have been
rare. The first LF studies covered the clusters M92 (Tayler 1954),
M3 (Sandage 1957), and M13 (Simoda \& Kimura 1968) down to about the
main sequence turnoffs. The typical precision of the photometry for
even the bright giants was such that the separation of the asymptotic
giant branch and first ascent giant branch was difficult, 
and the effects of blending and completeness on the observed LFs were
very difficult to
quantify with photographic data (although see Buonanno et al. 1994 for
a state-of-the-art photographic study of M3). 

Early CCD-based studies had the potential to increase the dynamic range
of LFs enormously, but because of the small field size of early
CCD cameras, the number of evolved stars was typically far too small
to be useful in testing evolution models. Attempts to mosaic small-field
CCD images and piece together LFs (Bergbusch 1993, Bolte 1994), meshing
photographic- and CCD-based studies (Bergbusch \& VandenBerg 1992;
hereafter BV92), and
deriving composite LFs combining the results from CCD studies of
several clusters
(Stetson 1991) have produced some intriguing results (see discussion below).
However, there are caveats that accompany all of these techniques, and the
results require verification.
  
The wide-field CCDs now in use can cover much larger areas
than the first generation of CCDs, so that for the first time it
is possible to derive accurate LFs for large
samples of stars in all phases of evolution. It is still not a trivial
task to measure these LFs because of
the computing requirements and various large and subtle effects
resulting from extreme crowding in typical GGC,
but the way is clear for LF investigations of stars in a
large number of GGC.

In addition to the ``classical'' tests of stellar evolution,
two unexpected observations of the past few years
require further investigation.
In a LF formed from the combination of the CCD-based LFs of the three
metal-poor clusters M68 (NGC 4590 = C1236-264), NGC 6397 (C1736-536),
and M92 (NGC 6341 = C1715+432), Stetson (1991) found an excess
of stars on the subgiant branch (SGB) just above the main-sequence
turnoff (MSTO). Bolte (1994) observed the metal-poor
cluster M30 (NGC 7099 = C2137-174) using a mosaic of small-field CCD
images and found a similar excess of SGB stars.
This excess is intriguing as it could be the observable result
of an unusually extended isothermal core in MSTO stars,
as would be produced by the actions of ``WIMPs'' (Faulkner \& Swenson 1993).
(The SGB is often defined differently by different
authors. We will take it to be the transitional region between the
main sequence turnoff and the base of the red giant branch
at the point of maximum curvature.)

Another unexpected observation involving the LFs is the mismatch
between theoretical predictions and the observed size of the ``jump''
dividing the
main sequence (MS) and the red giant branch (RGB). When a theoretical LF is
normalized to the MS, there is an excess of giants observed relative
to MS stars (Stetson 1991, BV92,
Bolte 1994). These results might be explained by the action of core
rotation (Larson, VandenBerg \& DePropris 1995). Because of the
potential importance of non-standard physics in stars, and because
of the caveats associated with earlier LF studies, the most
productive next step is to derive better LFs in a number of GGC.

M5 is ideal for these types of investigations. It is one of the
most massive clusters having moderate central density ($\log (M /
M_{\odot}) = 5.6$ and $\log \ (\rho_{0} / (M_{\odot} /
\mbox{pc}^{3})) = 4.0$; Pryor \& Meylan
1993), is fairly close ($\sim 8$ kpc; Peterson 1993), and is at
high galactic latitude ($b$ = 46\Deg8). Thus, it is
possible to measure a large sample of post-MS stars with minimal
field star contamination.
Because the well-studied globular cluster M3 (NGC 5272 = C1339+286;
Buonanno et al. 1994,
and references therein) has
physical characteristics that very much resemble those of M5, a
comparison of the two data sets should be quite valuable. In addition,
M5 is part of a set of clusters (with NGC 288 = C0050-268 and NGC 362
= C0100-711) showcasing the
``second parameter'' effect on HB morphology. One additional fact makes M5
unusual: according to Cudworth \& Hanson (1993), M5 has the largest
space velocity of the globular clusters studied to date, with a very
eccentric orbit that takes it far from the plane of the galaxy. What
impact its orbit may have on the intrinsic properties of the cluster is not
clear.

In the next section, we describe our observations of the cluster. In
\S 3, we describe the reduction and calibration of the program frames.
In \S 4, we discuss the features observed in the color-magnitude diagram.
In \S 5, we present the results of artificial star experiments that
were executed to 
determine incompleteness in the sample, and present the LFs. Finally,
in \S 6, we discuss the constraints that can be put on the global
parameters of the cluster --- metallicity, distance, and age.

\section{Observations}

The data used in deriving the $B$- and $I$-band LFs of M5 were taken on
June 14/15, 1993 at the Cerro Tololo Inter-American Observatory (CTIO)
4 m telescope.
In all, five exposures of 120 s and one exposure of 10 s were made in
$B$, and six exposures of 60 s and one exposure of 6 s were made in $I$.
All frames were taken using a 2048 $\times$ 2048
pixel ``Tek \#3'' CCD chip having a sampling of about 0\Sec48 per
pixel, and a field 16\Min3 on a side.  The
exposure times were chosen so that stars were observed on all of the
frames if they had magnitudes placing them
on the horizontal branch (HB), or on the principal sequence, fainter
than the level of the HB and brighter than about two magnitudes
below the turnoff. Our five
120 s $B$-band exposures were combined into one master $B$ frame, and the
six 60 s $I$-band exposures were averaged together to create two master
$I$-band frames, with the exposures being divided between a good-seeing
and a bad-seeing set. These longer exposures
were averaged together to reduce the effects of background noise. The two short
exposures (one $B$ and one $I$) were kept separate. Stars that
were saturated on the long exposures
were only measured on the short exposure frames, and stars up to two
magnitudes brighter than the detection limit were
only measured on the long exposure frames. All of the frames were centered
approximately on the cluster center, giving us a total radial coverage
of about $8^{\prime}$.

The night of these 4 m observations was not photometric. In order to
set the observations on a standard photometric system, we
used observations made at the CTIO 0.9~m telescope on two photometric
nights (June 16/17 and 19/20, 1993). The CCD used was the ``Tek \#1''
1024 $\times$ 1024 chip, having a size of 0\Sec396
per pixel, for a total sky coverage of
6\Min75 on a side. Landolt (1992) and Graham (1982) standard
star observations were used to calibrate a secondary field that
overlapped the 4 m field. The secondary field corresponds to the M5
West region used by Stetson \& Harris (1988) to calibrate their M92
data. On the two photometric nights, a total of 4 $B$, 7 $V$, and 8
$I$ exposures were taken. A sample of
118 stars having $13.0 < V < 18.5$ and $-0.06 < (B - V) < 1.26$ was 
calibrated as secondary standards in this way.

Supplementary observations of M5 were made in $V$
band on July 7/8, 1994 on the CTIO 4 m telescope. Three exposures of
10 s and three exposures of 60 s were taken. Again, the
shortest exposures were left separate, while
the longer exposures were combined into one
frame. The ``Tek \#4'' 2048 $\times$ 2048 CCD
chip was used during these observations. The pixels had a sky coverage
of 0\Sec44 each. In addition to the smaller field of view
for these observations, the frame center was offset west from the center of
the previous CTIO 4m observations by approximately
2\Min3. 

\section{Data Reduction}

\subsection{Primary Standard Calibration Fields}

All of the primary standard observations were made using the CTIO 0.9
m telescope, and only the first and last of those four nights of
observing were photometric, as determined from later analysis of
aperture photometry results. For the present discussion, we will
concentrate on issues related to the reduction of photometric data
that will be used to calibrate the 4 m observations.

For the 0.9~m frames, the bias level and pattern was removed by
subtracting both a fit to the overscan region and a master zero-level
frame. We found that the twilight flats removed the
remaining trends on the frames (especially low-frequency variations)
in the best manner, and so they were used exclusively in flat-fielding
the object frames.

\subsubsection{Aperture Photometry}

Primary standard stars were observed on the two photometric nights and
on one nonphotometric night (June 18/19, 1993). The standard star
observations were taken at a
range of airmasses spaced throughout the nights in order to determine
the extinction coefficient, and the fields
themselves were chosen to give good color coverage.
Data from the nonphotometric night was only used to help constrain the
color terms in the transformation equations. Aperture photometry was
performed using the program DAOPHOT II (Stetson 1987). 
Using the aperture photometry data, growth curves were constructed
for each frame, in order to extrapolate from the flux measurements
over a circular area of finite radius to
to the total flux observable for the star.
For this purpose, the program DAOGROW (Stetson 1990) was employed.

The aperture magnitudes and the known standard
system magnitudes of Landolt (1992) and Graham (1982), with some
updated values from Stetson \& Harris (1988), were then used to
derive coefficients of the transformation equations, which are as follows:
\[ b = B + a_{0} + a_{1} \cdot (B - V) + a_{2} \cdot (X - 1.25) +
a_{3} \cdot t \]
\[ v = V + b_{0} + b_{1} \cdot (B - V) + b_{2} \cdot (X - 1.25) + b_{3}
\cdot t \]
\[ i = I + c_{0} + c_{1} \cdot (V - I) + c_{2} \cdot (X - 1.25) + c_{3}
\cdot t ,\]
where $b$, $v$, and $i$ are observed aperture photometry magnitudes, $B$, $V$,
and $I$ are the standard system magnitudes, $X$ is airmass, and $t$ is
Universal Time (UT) of observation. The time terms were only added after an
initial calibration using the zero point, color, and airmass terms
showed a very slight trend in the residuals with this variable.
As the strength of a color term is a partial reflection of the mismatch between
the transmission curve of the filter used and that
of the standard system filter, a $(V - I)$ term
was used in the $I$-band transformation equation, instead of a $(B - V)$
term. The primary standards used in the calibration covered a color
range $-0.3 < (B - V) < 1.3$.
The color-dependent term was also determined on the
nonphotometric (third) night by a similar method, with the photometric zero
point of each frame allowed to vary independently of the others, so as to
remove the effects of variable extinction during the night. As a
result, neither airmass nor time terms were used in the transformation
equations for that night.

To improve the evaluation
of the color terms (which should be fairly consistent from night to night),
some measurements were eliminated. The stars 110-502 and T Phe D
(Landolt 1992) were
removed due to their extremely red 
colors. As these were the only stars observed with $(B-V) > 1.5$, their
inclusion would have required several additional color terms to model.
The largest effect of this choice is on $B$ band, which matches
the standard filter transmission most poorly. Higher order
color terms would have been necessary to effectively calibrate this
band because of the rapidly falling intensity distribution from red
stars in this wavelength band --- the measured $B$-band magnitude would be
abnormally sensitive to temperature changes. With the linear
term used here we get consistent values from night to
night. The systematic nature of magnitude errors for $B$-band
measurements for red stars should be kept in mind though --- stars with
$(B - V) > 1.5$ could be measured too faint relative to standard
values by approximately 0.04 mag, if we can judge from the
residuals of 110-502 and T Phe D.

The color terms derived for the three nights on which standard stars
were observed were then combined in an average weighted by the
calculated errors in the terms. These color terms are presented in
Table~1a, along with the error in the mean. The derived values for the
color coefficients were fixed in
a redetermination of the coefficients of the other terms for the
photometric nights. The values of these
coefficients are shown in Table~1b. The time term was only found to be
significant on the first night.

While it has been found that there have been reports of shutter delays
of several tens of milliseconds on the telescopes at CTIO, none of our
exposure times on the 4 m or 0.9 m telescopes were shorter than 6
s. As a result, the shutter corrections are 0.005 mag at most. We have
opted not to make any attempt to compensate for this.

With the coefficients determined, the observed magnitudes of the
standard stars were run through the transformation equations, and
multiple measurements were combined to get a file of observed values
of the primary standard magnitudes. A comparison with the published
values was made to determine the average residuals. (In this
and all subsequent comparisons, the
residuals are calculated in the sense of ours -- theirs.) The average
residuals for these stars are expected to be
low; the comparison for the sample of 38 stars is shown in
Figure~\ref{fig1}, and the average residuals are given in Table~3.

\subsection{Secondary Standard Calibration}

The M5 secondary standard stars were calibrated from the
observations on June 16/17 and 19/20 via a method similar to that used
for the primary standards. The unsaturated stars found in all frames
with magnitude placing them above the turnoff were included in the
first list. Stars which appeared to fall in the RR Lyrae gap were
excised, as were stars that fell too far from the fiducial lines of
the cluster, and stars that had neighbors within about 10 pixels.
Three blue horizontal branch stars were returned to the
list even though they were not present on all frames, so as to improve
the color coverage of the sample. Once the list was finalized, all
other stars were subtracted from the frames. 

Aperture photometry was taken for the stars on these secondary
standard frames. Repeated observations were combined with
weights based on the square of the measurement error in order to create
the final library of magnitude values. The values for a total of 118
stars are provided in
Table~2, along with cross identifications to M5 calibration standards
in Stetson \& Harris (1988). The field is approximately $9^{\prime}$ west of
the cluster center --- a finding chart is provided in
Figure~\ref{fig2}. A comparison of values for 34 stars is shown in
Figure~\ref{fig3}, and the average residuals are given in Table~3. Values from
Stetson \& Harris (1988) were not used in the
comparison, however, as it was found that there was a trend in
residuals with Y coordinate (approximately 0.07 mag in B over 500
pixels, and 0.04 mag in V). Newer values provided by Stetson
(1994b) did not show this effect. There is still a trend in the
residuals with color that would account for the difference in $V$
magnitudes. This may be the result of our smaller primary standard
color range compared to Stetson \& Harris. Since our sample
is optimized for the range of color in which all of the comparison
stars fall, we will continue to use our library of secondary standards to
calibrate the CTIO 4 m data.

The secondary standard field covers the majority of the faint star
field from Arp (1962), and to a lesser extent the middle field from
Richer \& Fahlman (1987; hereafter RF). The secondary standards that
RF calibrated were fainter than any we used, so no comparison will be
made here.

\subsection{Object Frames}\label{3.3.}

\subsubsection{Profile Fitting Photometry}\label{psf}

The CTIO 4 m data were reduced using the standard suite of programs
developed by Peter Stetson: DAOPHOT/ALLSTAR/ALLFRAME (Stetson 1987,
1989b, 1994a). Stars were
chosen on each of the master $B$, $V$, and $I$ frames in order to determine the
point-spread function (PSF), and its variation with position on the
frame. Each frame was divided into 25 bins, and the 15 brightest stars
from each subsection were put into a list. The average FWHM for these
stars was calculated, and stars which deviated from this value by more
than $3\sigma$ were discarded. The radial profiles of the stars were
then viewed to beyond the radius to be included in the PSF to
eliminate stars with contaminants, such as faint companions. Typically
about 150 stars were used in determining the PSF and
its spatial variation on each frame. There was a definite bias against crowded
portions of the field in the selection of PSF stars. However, because
the frames were centered on the cluster, PSF variations across the
frame were mapped well.

Three passes were made using a combination of DAOPHOT's FIND routine
and ALLSTAR to derive a star list for each frame. These star lists
were combined according to filter of observation, and then the filter master
lists were combined into a master star list. Stars were only kept if
they were detected on at least one frame in each filter. The master
list of stars, and the frames and their PSFs were supplied as input
into ALLFRAME (Stetson 1994a), which simultaneously reduces stars on
all of the frames at once in order to derive consistent positions.

The correcting lens used in taking the $V$ frames on the
CTIO 4 meter telescope had the
effect of modifying the coordinate system of the $V$-band frames. While
linear coordinate transformations did a reasonable job in mapping
between frames from the two runs, there was a scale change in the
sky coverage of the pixels with distance from the center of the chip.
This should not affect the photometry (as the PSF variations should
easily take this into account), but we resorted to running ALLFRAME on
the $V$ frames separately. The $BVI$ overlap sample was derived by
matching the $V$ and $BI$ samples together.

\subsubsection{Calibration}\label{cal}

Two different samples were created from the $B$, $V$, and $I$ frames, and
each was calibrated separately. Because the $B$ and $I$ frames were better
centered and had better seeing, the sample was larger by about 25\%. 
54 of the secondary standards observed in the CTIO 0.9 m sample also
fell on these 4 m frames. The transformation equations used in the
calibration were:
\[ b = B + a_{0,k} + a_{1} \cdot (B - I) + a_{2} \cdot (B - I)^{2} \]
\[ i = I + c_{0,k} + c_{2} \cdot (B - I)^{2} .\]
The coefficients of the color terms in this case were $a_{1} = -0.0349$,
$a_{2} = 0.0349$, and $c_{2} = -0.0095$. A weakness of this
calibration is the fact
that only one extremely blue ($(B - I) < 0$) and three extremely
red stars ($(B - I) > 2$) fell in the section of overlap.

The $V$-band frames from the 1994 run were taken with the cluster off
center, resulting in a smaller overlap with the earlier $B$ and $I$
frames. So, a smaller sample consisting of all stars detected in $B$, $V$,
and $I$ was also calibrated using the equations: 
\[ b = B + a_{0,k} + a_{1} \cdot (B - V) + a_{2} \cdot (B - V)^{2} \]
\[ v = V + b_{0,k} + b_{1} \cdot (B - V) + b_{2} \cdot (B - V)^{2} \]
\[ i = I + c_{0,k} + c_{2} \cdot (V - I)^{2} .\]
The color terms are shown in Table~1c.

The size of the CTIO 4 m frames means that most other surveys of M5
overlap the program area at least partially. Table~3 provides a
summary of several of the zero-point offsets for comparisons with
several of these studies. The field of
the Buonanno, Corsi, \& Fusi Pecci (1981; hereafter BCF81) survey is
completely included on all 4 m frames.
We clearly see scale differences (a systematic
trend in the residuals with magnitude),
color-dependent residuals, and zero-point differences between our
photometry and their $B$- and $V$-band data of BCF81, as is typical for the
other photographic surveys.

The frames also completely enclose the inner field of RF.
They tabulate values for 12 stars in common with BCF81, and for the
7 secondary standards they used in this field.
There does seem to be a small systematic trend of
$\Delta (B-V)$ with $(B-V)$ across the entire range of color in our
comparison with RF.
We have also made a larger comparison with the photometry of Storm,
Carney \& Beck (1991; hereafter SCB) for a
region corresponding to field IV of Arp and BCF81, as shown in
Figure~\ref{fig4}. We have 1584 stars in common, but we only show
comparisons for the 384 stars with $V > 17.5$ as fainter stars in SCB's
photometry have much larger uncertainties. A trend in the color
residuals with color also seems to be
present, in the same sense as in the RF comparison. We are inclined to
believe our results due to more extensive primary and secondary calibrations.

Calibrated $I$-band data for the cluster is rather sparse. The best
source of comparison is with the photoelectric photometry of Lloyd
Evans (1983). Figure~\ref{fig5} shows the comparison with his sample
of 22 stars from the Arp (1962) sample, with two additional stars near
the red-giant tip. This comparison is of some importance, as the
position of the upper RGB in $V$ and $I$ band is often used as a
metallicity indicator (Da Costa \& Armandroff 1990; Sarajedini 1994).
There does appear to be systematic zero point offsets in $V$ and $I$
bands of approximately 0.02 mag, in the sense that our values are
brighter. (It should be noted that there is good agreement between the
average zero-point offsets in $V$ band for the comparisons with RF,
SCB, and Lloyd Evans (1983) data.) The zero-point differences are
small enough that they should not present a problem in using the upper
RGB to determine the metallicity.  The star TLE1, which is near the
tip of the RGB, has the largest absolute residual of any star in the
comparison sample. The large residual for this star may be partly the
result of our calibrations procedure, as the color of the star $(V =
12.056, (B-V) = 1.653, (V-I) = 1.746)$ puts it outside of the range of
colors over which the calibration was optimized. However, it is also
the star of Lloyd Evans' sample that is in the most crowded region,
and would be most likely to be affected by background considerations.
The agreement with the photoelectric value is reassuring.

\section{The Color-Magnitude Diagram}\label{4.}

Table~4 presents the fiducial lines determined from the CMD of the
cluster. We also include the number of stars used in the calculation of the
fiducial line for each bin. Due to the poorer quality of the $V$-band data,
especial care had to be paid to ensure that the effects of blends and crowding
did not overly influence the values derived. Fiducial points for the
MS and lower RGB were determined by
finding the mode of the color distribution of the points in magnitude
bins. Blends of stars produce significant redward
biases in the fiducial determination, particularly on the MS,
so that the use of the mode is preferred over a mean. A cut was made
on projected radius ($r > 350$ pix = 2\Min8) to restrict the
sample to the best measured stars. The position of
the SGB was determined by finding the mode of the star distribution in
color bins. This was done because the SGB is close to horizontal in the
CMD, so that mode finding in magnitude bins was affected by the blend
sequence.

The fiducial line on the RGB was determined by finding the mean color of
the stars in magnitude bins. Once a mean was determined, stars falling
more than $5\sigma$ from the fiducial point were discarded (so as to
eliminate AGB and HB stars, as well as blends and poorly measured
stars), and the mean redetermined. This procedure was iterated until
the star list did not change between iterations. The lower portion of
the AGB was also measured in this way. At the tips of the RGB and AGB,
the positions of individual stars were included as fiducial points if
they appeared to be continuations of the mean fiducial line. The
fiducial line for the HB
was determined by determining mean points in magnitude bins for the blue tail,
and in color bins for the horizontal part of the branch. It should be
stressed that the HB fiducial points
represent the most populated portion of the HB, and not the zero-age line.
For the RGB, AGB and HB, a color error cut ($\sigma_{B-V} < 0.015$)
was used to restrict the
sample to the best measured stars. All sequences were checked for
continuity in regions of the CMD where 
fiducial points could be derived by different methods. No smoothing
has been applied.

Figure~\ref{fig6} shows a comparison between the fiducial sequence of the
present data set and those of BCF81 and RF. Also included is BCF81's
determination of the fiducial lines of the RGB and AGB from Simoda \&
Tanikawa's (1970) data. There is fairly good agreement among the
data sets. For example, our fiducial points fall directly between the
RF inner and outer field values at the turnoff, and also match the SGB
quite well. Our fiducial is bluer than either set of RF points lower
on the MS by as
much as 0.04 mag. This is probably not surprising, as RF determined
their fiducial points by taking means in magnitude bins. The mean
would tend to be redder than the mode on the main sequence due to the
inclusion of unresolved blends or binaries. However, our fiducial
points for the lower RGB also lie to the blue of the RF points. The
upper RGB fiducial points show some systematic disagreement with those
derived from the photographic studies, but this is not significant in
light of the systematic trends known to exist in the photometry from
those studies (see \S \ref{cal}). The AGB points agree very well with those
of BCF81.

The $BI$ fiducial was easier to determine due to the larger sample,
better seeing for the frames (which improved the photometry in the
cluster center), and the better sensitivity of the $(B-I)$ color to
surface temperature differences (color errors are smaller relative to
the total color range, compared to $(B-V)$). Only a radial cut was
applied ($r > 200$ pix = 1\Min6). Mean RGB points are also tabulated for
$(I,V-I)$ in order to allow us to compare the position of the giant
branch with those of other clusters (Da Costa \& Armandroff 1990).

In Figure~\ref{fig7}, we plot the total samples for both the $BVI$ (28339
objects) and $BI$ (42456 objects) sample. It is apparent that the
quality of the $(I,B-I)$ data is quite superior to the
$(V,B-V)$ data. Both
samples cover the center of the cluster. Several conclusions can
be gleaned from the $(I,B-I)$ CMD. First, it should be possible
to make distinctions between stars on different evolutionary
branches once the most crowded regions are removed. Second, there is a fair
number of galaxies to be found on the frames. This is the reason
for the population seen well to the red of the MS. Third, there
is a hint of a blue straggler star sequence between the MS turnoff
and the blue end of the HB (which becomes obvious for more
restrictive cuts on the sample).

Figure~\ref{fig8} presents plots of restricted subsets of the data
plotted in Figure~\ref{fig7}. The total $BVI$ and $BI$ samples were
sorted according to CHI value (CHI is related to the quality of the
object image; see Stetson 1989a). Less restrictive cuts were used for
the bright stars (above the SGB) in order to maintain the definition
of the branches in the CMD. In the ($V,B-V$) diagram, for stars
with $V < 17.4$ only stars with CHI $<$ 1.3 were kept, whereas for $V
> 17.4$, only stars with CHI $<$ 1.07 were kept. In the ($I,B-I$)
diagram, CHI $<$ 1.6 for $I < 16.5$, and CHI $<$ 1.09 for $I > 16.5$.

\section{Determination of the LF}\label{makelf}

\subsection{Artificial Star Tests}\label{5.1.}

A particularly crucial component of the calculation of an accurate LF
is the determination of incompleteness corrections. For the data at
hand, crowding is the primary source of incompleteness. Because of the
large range of crowding conditions present in the observations, it is
necessary to determine these corrections as a function of position
(projected radius) as well as magnitude. To this end, we
carried out extensive artificial star tests to correct the data for
biases of various sorts.

The bias-subtracted and flat-fielded $B$- and $I$-band frames were the
only ones used in this portion of the data reduction (although the
long exposure $I$ frame with poorer seeing was discarded in
order to reduce the amount of computer time used in the calculations).
The PSFs we used were simply those determined by the procedure
outlined in \S \ref{psf}. The output of the reduction of these frames
using DAOPHOT/ALLSTAR/ALLFRAME without artificial stars is considered the
``control'' run. A
virtually identical procedure was followed in doing the artificial
star experiments. The starting data were: the $B$ and $I$ frames, the PSFs
for each frame, the fiducial lines of the cluster and an initial
luminosity function. A
theoretical luminosity function was initially used to give the
probability that an artificial star would have a particular
magnitude. The set of artificial stars was thus weighted toward the
faint end. Initially, artificial stars were only allowed to be
brighter than about two magnitudes below the turnoff in $I$ band, so as
to allow the program to create an adequate number of artificial red
giant stars for analysis.
This procedure has the advantage of allowing us to
mimic the LF of the cluster so that we can get a fair estimate of the
amount of photometric scatter created by blending. 

The fiducial lines of the cluster were determined in instrumental magnitudes
from the control run. Once a magnitude was picked in one band
($I$-band in our runs), the magnitude in the other band was chosen so as to put
the artificial star on the fiducial. Positions were chosen randomly,
so that they had an even distribution on the frame. Each star was
given a
consistent position on all frames for which the star fell in the field of view.
Given the previously determined form of the PSFs for
each of the frames, the artificial star images were placed on the frame
using the DAOPHOT ADDSTAR routine. The resulting frames were put
through a reduction procedure identical to that for the control run
(starting after the PSF determination) to
ensure that the artificial stars were treated exactly like the real stars.
No more than 3000 artificial stars were put on the frames at any one
time so that the artificial stars did not significantly affect the
completeness at any radius. In 20 artificial star runs,
a total 48,058 artificial stars were added and reduced.

The output from each artificial star run was a list of stellar
positions and magnitudes for all stars that were classified as
``detected.'' It is not a trivial matter to determine which
detected stars should be identified as input artificial stars.
Faint artificial stars that happened to be placed on bright real stars
would not be detected as such. If you simply look at the positions at which
you know artificial stars were placed, your search would
identify the bright real star as the artificial star. If you
decide to discard the star in the above example, you would be
throwing away data on the extreme limit of the blending problem --- how
the photometry of a cluster of stars is affected by the imperfect
resolution of the instrument. 

In keeping with this philosophy, a simple positional search was
conducted on the artificial star run photometry, and the input
magnitudes were recorded, along with the final magnitudes if any object
was found at the input position (to within a pixel). 
Limited computer time forces us to try to decouple the effects of
blending and crowding, even though they are clearly linked to
increases in stellar image density on images. Crowding effects can be
corrected by determining sample completeness versus magnitude
and projected radius. A correct simulation of blending requires adding
stars several magnitudes fainter than the range of magnitudes in which we are
interested. If a real star is blended with the faintest
artificial star we added, the resultant star will be brighter by a
certain magnitude increment, which will be nonzero. In practice, the
fainter the added star the more negligible the increment.

It is then necessary to reanalyze the artificial star sample so that
artificial stars that were blended with much brighter stars are
labeled as undetected and so will count toward incompleteness. One
pass was made through the artificial star data to make preliminary
calculations of the external errors in the photometry $\sigma_{ext}$
by looking at the scatter in the differences between the input
and output magnitudes. A second pass was then made, and any artificial
star that was detected more than 1.0 magnitude brighter than its input
magnitude or more than $5 \sigma_{ext}$ from the fiducial was labeled
as undetected. The resulting sample of recovered artificial stars was
then used to
calculate the following quantities in bins sorted by projected radius and 
magnitude: 1) median magnitude ($B$ or $I$), 2) median color $(B-I)$,
3) median internal error estimates ($\sigma_{B}$ or $\sigma_{I}$, and
$\sigma_{B-I}$), 4) median magnitude and color biases ($\delta_{B}
\equiv \mbox{median}(B_{\mbox{output}} - B_{\mbox{input}})$ or
$\delta_{I}$, and $\delta_{B-I}$), 5) median external error
estimates ($\sigma_{\mbox{ext}}(B) \equiv \mbox{median}|\delta_{B} -
\mbox{median}(\delta_{B})| / 0.6745$ or $\sigma_{ext}(I)$, and
$\sigma_{ext}(B-I)$), and 6) total recovery probabilities ($F(B)$
or $F(I)$ --- the fraction of stars added to the bin that were recovered
with any magnitude). The width of each magnitude bin was adjusted so as to
put equal numbers of recovered artificial stars in
each, in order to try to maximize the accuracy of the median for quantities 1)
- 5). The binning for the total recovery probabilities was done
somewhat differently -- bins containing equal numbers of recovered
stars were used along with the unrecovered stars in the bin to give $F$.
This method loses sensitivity near the detection limit. To
characterize this region, bins containing equal numbers of unrecovered
stars were formed, and the number of recovered stars in the bin was
used to give $F$. For magnitudes for which larger numbers of artificial
stars were used, several bins were combined so as to reduce the noise
in the determination of the incompleteness quantities.
The result was very well-determined curves for $F$ in the different
radial bins.

The runs of these quantities naturally show noise, and do not
cover the entire range of detected real stars. In order to smooth out
noise and to enable reasonably accurate extrapolation beyond the
magnitude range of the artificial stars, we looked for functions that
would fit the data. The functional forms chosen for accuracy and
stability were:
1) magnitude and color error estimates:
\[ \sigma(B) = \sigma_{0}(B) + \exp[a_{1} + a_{2} B] \]
The magnitude errors are well fit by a simple exponential --- the errors
go up substantially at the completeness limit as the typical size of
the noise peaks becomes comparable to the peak size of the stars under
consideration.  
2) magnitude biases:
\[ \delta (B) = \delta_{0}(B) + b_{3} B - \exp[b_{1} + b_{2} B] \]
The magnitude bias tends to become more negative with increasing magnitude in
crowded regions and at the completeness limit. In other words,
the artificial star is measured brighter, whether due to
contamination from the light of other stars,
or due to noise spikes that create a detection bias for stars that
fall on positive spikes. The linear term was included to take account
of low-level trends with magnitude seen in the outermost radial bins.
These trends are not visible in plots in similar studies (Figures 21
and 41 of Stetson \& Harris 1988, Figure 16 of Bergbusch 1993) due to
scale. These trends only amount to 0.002 magnitudes of bias per
magnitude, so it was not necessary to correct for this in the CMD. At the
completeness limit there is real scatter in the median points.
3) recovery probabilities ($0 \leq F \leq 1$) :
\[ F(B) = 1 - \exp[c_{1} + c_{2} B] 
- \left\{ \begin{array}[c]{ll}
               0   &  \mbox{if $B < B_{crowd}$} \\
               (c_{3} + c_{4} B)   &  \mbox{if $B \geq B_{crowd}$} 
          \end{array} \right. \]
where $B_{crowd} = -c_{3} / c_{4}$. 
There are three magnitude regimes inherent
in this formulation. The first is the complete sample -- for the
brightest stars in the cluster, there are not enough stars of equal or
greater brightness to allow them to get lost. The linear portion of the curve
models the effects of crowding (see Figure 3 of Stetson 1991), and
essentially just reflects the growing number of stars bright enough to
create blends with a star of the given magnitude and significantly
modify the measured brightness. Finally, the
exponential cutoff fits the detection limit of the frames, probably
reflecting the fact that noise spikes are less and less likely to
be able to move a given faint star above the detection threshold.
In our sample, the linear regime begins approximately at
the cluster turnoff for the outermost radial bins, due to the rapidly
rising number of stars. It begins brighter in the innermost radial bin
due to the background of unresolved stars. This represents the
practical limit to our ability to find all of the stars in the
frame of a given brightness under the seeing conditions in the
atmosphere and with the resolution capabilities
of the telescope.

Errors on each of these quantities were estimated using a similar
procedure. The points in each radial bin were adjusted upwards and
downwards by a fraction of the errors on the points, and the
adjusted points were fitted using the same functional forms. The
fractional adjustment was chosen so that the adjusted curves
encompassed the majority of the points between them. The constraint
that $F$ approach 1 for bright stars was retained in all cases, with the
result that the error in $F$ at the bright end is effectively zero. The
results of these calculations for $2^{\prime}$ radial bins are shown in Figures~\ref{fig9}-\ref{fig12}.

The data derived from the artificial star tests were used to derive
corrections to the observed luminosity function according to the
procedure outlined in Bergbusch (1993), which was based on Stetson \&
Harris (1988), and ultimately on Lucy (1974). This method uses the
error distribution, magnitude bias, and recovery probability to
predict the form of the
observed luminosity function $\eta(m)$, given an initial guess for the ``true''
luminosity function $\phi(m_{t})$:
\[ \eta(m) = \int_{-\infty}^{+\infty} \phi(m_{t}) P\{m \mid m_{t}\} dm_{t}
\]
where $P\{m \mid m_{t}\}$ is the probability that a star of true magnitude
$m_{t}$ is measured at magnitude $m$. Following Stetson \& Harris, we
will assume
that this probability distribution has the form
\[ P\{m \mid m_{t}\} = \frac{F(m_{t})}{\sqrt{2\pi}\sigma(m_{t})}
\exp\left(\frac{-[m - m_{t} - \delta(m_{t})]^{2}}{2\sigma^{2}(m_{t})}
\right), \] 
or in other words, a set of stars of true magnitude $m_{t}$ will be
measured with a Gaussian distribution around the biased magnitude
$m_{t} + \delta(m_{t})$. The initial guess for the true LF can be
corrected iteratively by comparing the observed LF to the predicted LF
and adjusting the bins that contribute to discrepant predicted
bins according to 
\[ \phi^{r+1}(m_{t}) = \frac{\int^{+\infty}_{-\infty} [\tilde{\eta}(m)
/ \eta(m)] \phi^{r}(m_{t}) P\{m \mid m_{t}\}
dm}{\int^{+\infty}_{-\infty} P\{m \mid m_{t}\} dm} \]
where $\tilde{\eta}(m)$ is the observed LF. This procedure matches the
observed LF best in just a few iterations, even though a calculation
indicates that $\chi^{2}$ is still decreasing (Lucy 1974).

A few comments should be made about the usefulness of this procedure.
Clearly it is designed to account for magnitude biases that would
cause stars to move in magnitude from bin to bin. This effect is
especially important in regions where crowding effects are important
and near the detection limit, as can be seen from Figure~\ref{fig7}. 
This procedure, however, cannot correct for blending effects
explicitly. This can be done most simply and effectively by
restricting the sample to those
regions of the cluster in which the effect is least important ---
uncrowded regions in the outer parts.

Once this procedure is completed, the completeness fraction $f$ can be
calculated simply by taking the ratio of the predicted number of
observed stars to the true number of stars (taken from the derived
``true'' LF). The calculated values of the completeness fraction were
also fitted using the functional form used for the total recovery
probability $F$. The value of $f$ for a star of any magnitude can then
be calculated by interpolation in the values at various radii.
To calculate the total LF, the real observed stars can just be read
in (multiplied by a factor of $f^{-1}$ to account for incompleteness)
and binned. No star was included if the completeness factor f was less
than 0.3. This condition defines a projected radius cut-off for each
magnitude bin.

One additional step was then taken to try to increase the accuracy of
the LF values for the RGB, where Poisson statistical noise dominates
over incompleteness. For the faintest magnitude bin
considered, we cover the minimum amount of cluster area fairly
completely ($\overline{f} > 0.3$). 
For the brightest stars in the sample, we are
essentially assured of detecting them anywhere on the frame --- even
into the core of the cluster. Because of this, we decided to count
all of the bright stars in the sample. Following Bolte (1994), we
scale the total number of
stars detected to the number detected within the area for which
the faintest bin in the LF was more than 30\% complete. In other
words, the corrected number
of stars in a bright bin is $N_{i}' = s N_{i}$, where $N_{i}$ is
total number of stars detected in the cluster that fell in
magnitude bin $i$. The normalization factor is
\[ s = \frac{\sum_{i=n_{1}}^{n_{2}} N_{i}(r >
r(n_{tot}))}{\sum_{i=n_{1}}^{n_{2}} N_{i}} ,\]
where $n_{tot}$ is the faintest bin in the LF. $r(n_{tot})$ is the
projected radius at which stars in the faintest bin are 30\% complete.
Bins in the range $n_{1} \leq i \leq n_{2}$ must be 100\% complete, or
already corrected for the incompleteness as a function of radius.
By definition, $s \leq 1.0$.

The scale factor
$s$ was determined for a magnitude range on the lower RGB, and
the value was applied to all of the magnitude bins included in
this bright sample. We found that the value of $s$ remained quite
constant on the lower RGB, but increased within a couple of
magnitudes of the RGB tip. This is an indication that there may be a
deficiency of giants on the upper RGB in the center of the cluster
(a larger fraction of the stars in those bins were from the
outskirts of the cluster). The constant value found on the lower
RGB is more likely to reflect a population that has not been
affected by dynamical processes in the cluster. By using this
value of $s$ for all bright magnitude bins, we are able to
get ``global'' values for the LF in these bright bins,
incorporating the deficiencies of RGB stars. The deficiency of
upper RGB stars in the core of the cluster is reflected in the
population of the innermost annulus in Table~7.

The LFs determined via the procedure outlined in \S \ref{makelf} are given in
Table~5. A total
of 22044 and 21388 detected stars were used in the calculation of the cluster's
true LFs in $B$ and $I$ bands respectively. The faint portion of the
sample was restricted to projected radii
greater than 450 pixels from the center of the cluster for $B > 17.8$
(and for $I > 16$ in its LF) in
order to diminish the effects of blends. The normalization regions
for the two LFs were $21 < B < 21.5$ and $19.6 < I < 20.1$. 

\subsection{Image Blending}\label{5.2.}

The effects of optical blending on the photometry of cluster stars
must also be properly gauged and eliminated in order to ensure the validity
of the star counts. The effects of
blending are quite obvious in CMDs of dense regions of globular
clusters. Blends involving MS or RGB stars can be seen in a swath
bounded on the faint side by the fiducial line of the
cluster, and on the bright side by the same fiducial offset by about 0.75
mag. Blends are most important on the main sequence and subgiant
branch because of the sheer numbers of stars that can potentially be
blended --- if one of the stars being blended is too faint relative to
the other, it will not change the luminosity of the blended star far
from that of the brighter component. On the main sequence, blended
stars can potentially be mistaken for binary stars. On the subgiant
branch, they can add erroneously to brighter bins. Figure~\ref{fig15} shows the
full effect of blending on the $B$-band LF when no restrictions are
made.

In the turnoff region and at the base of the RGB, the blend problem is
especially insidious because two
stars of approximately the same brightness create a blended star that
falls very near the cluster fiducial because the fiducial is vertical
at that point in the CMD. Such stars will then pass any tests
that reject stars based on distance from the cluster fiducial. An
additional effect is created by the horizontal nature of the SGB in
$B$-band. Blended SGB stars move more nearly
perpendicular to the fiducial line of the cluster, and are therefore
much more likely to be eliminated by cuts based on proximity to
the fiducial. In the case shown in Figure~\ref{fig15}, when crowded
regions are included in the sample, the LF can appear to reflect that
of a more metal-poor or older cluster (see discussion in \S \ref{isolf}).

One can potentially use Monte Carlo simulations to derive
statistical corrections for blending. However, the simplest method of
reducing the problem is to measure large samples of stars in
relatively sparse regions of a massive cluster. In 
this way, the effect of blending can be minimized, and a comparison of
the LF derived at large radii with an LF including crowded regions
gives an estimate of the size of the effect.

The choice of filter band can also change the impact of blended stars
on the LF. In $V$ band, the subgiant branch has a relatively flat
slope in the CMD. As a result, the entire subgiant branch falls within
0.75 mag of the turnoff, and so is especially susceptible to
blending. In the survey at hand, we have used both $B$ and $I$ magnitudes 
to construct luminosity functions. In $B$ band, the subgiant
branch covers a smaller range of magnitudes than in $V$ band, and
so the LF will have 
the same problems as a $V$ band LF. In $I$ band, the SGB has a much
steeper slope in the CMD. As a result, the blend sequence will
make contributions over a wider range of magnitudes, and
will be diluted in proportion.

Careful attention was paid to the upper RGB in an attempt to push the
LF as bright as possible. It is difficult to distinguish between
individual AGB
and RGB stars near the tip, although the fiducial sequences can be
followed. We found that our procedure for
eliminating outliers in the CMD was too conservative on the upper RGB.
To circumvent this, we varied the threshold value of the external
error used to make the cut until we found an optimal value that seemed
to keep RGB stars and eliminate AGB stars. There will undoubtedly be
contamination to an extent, and this will be the dominant source of
systematic error.

We will defer the remainder of the discussion of the LFs until \S \ref{lfdisc}.

\section{Discussion}\label{6.}

Because there are a large number of parameters (most importantly,
chemical composition, distance, and age) that influence the
photometry of a globular cluster, we now discuss the ways in which
each of these parameters can be constrained, and the expected reliability
of the results we derive. In the discussion that follows,
we will generally assume E$(B-V) = 0.03\pm0.01$.

\subsection{Metallicity}\label{met}

Sneden et al. (1992) provide a summary of many of the metallicity
determinations for M5 
made in the past 20 years. Among medium- and high-resolution spectral
analyses of giants using techniques similar to those by Sneden et al., the
[Fe/H] values range from --1.09 to --1.5. The authors found that they were able
to reconcile four out of the five studies with their measured value of
$<$[Fe/H]$> = -1.17 \pm 0.01$ from 13 giants. Sneden et al. also
briefly discuss four
determinations based on low-resolution spectrophotometric methods that
were calibrated using stars having high-resolution data. Three out of four of
these studies also returned metallicity values at the high end of the
range, with the Zinn \& West (1984; hereafter ZW84) study being at the
low-metal end. The ZW84 data were calibrated using high-resolution data from
Cohen (1983), which returned [Fe/H] = --1.4.

While the majority of the determinations are consistent, the absolute
zero points for the metallicity scales in the various investigations are
certainly offset relative to each other. This is an important issue
because the absolute metal content
does play a role in the choice of isochrones and also in our
determination of the distance modulus via
subdwarf fitting. Of the high-resolution spectroscopic determinations,
the Sneden et al. result is to be preferred in an absolute sense for
their use of improved
input physics and spectra. The results of the Lick-Texas group (see
Kraft et al. 1995 for the latest in this series) also form an
excellent relative abundance system. 

If the Lick-Texas group results are directly compared with ZW84
measurements, there is a noticeable nonlinearity - in the sense
that the metal-rich and metal-poor ends of the range are
systematically higher in the ZW84 scale, and systematically
lower at intermediate metallicities (around that of M5). With a sample
of 21 clusters having high dispersion spectroscopy, Carretta \&
Gratton (1996) find that a second order fit is preferred
to a linear fit. These studies provide additional reason for
considering metallicity scales beside that of ZW84. 
We will continue to regard the absolute metallicity of M5 as
uncertain at the 0.2 dex level, and we will attempt to propagate the
this range through the following analysis.

Several methods of checking the {\it relative} metallicity are open to
us using the
BVI data set. Three indicators are $\Delta V_{1.4}$ (Sandage \&
Wallerstein 1960), $(B - V)_{0,g}$ (Sandage \& Smith 1966), and RGB
position in $(M_{I}, (V - I)_{0})$ (Da Costa \& Armandroff 1990).

Figure~\ref{fig21} shows the CMD for the stellar sample used to calculate $(B -
V)_{0,g}$ and $\Delta V_{1.4}$. From the CMD, it is apparent that there
is a nonzero slope to the horizontal branch across the RR Lyrae strip.
From 20 stars nearest the blue edge of the
RR Lyrae strip we get $V_{HB,be} = 15.079 \pm 0.029$, and from 18 stars
near the red edge of the strip we get $V_{HB,re} = 15.105 \pm 0.021$.
From 11 RR Lyrae variables for which we have intensity-weighted mean $V$
magnitudes (SCB), we get $V_{RR} = 15.057 \pm 0.028$.
There appears to be a slight offset between our data and that of
SCB, so we will just adopt the value $V_{HB} = 15.092 \pm 0.017$
in order to remain consistent with the rest of our photometry.

From the fiducial line of the cluster, we can derive the colors of the
RGB at the level of the HB: $(B - V)_{g} = 0.855 \pm 0.002$ and $(V -
I)_{g} = 0.966 \pm 0.003$. The quoted errors reflect the error in the
determination of $V_{HB}$ and the slope of RGB. The adopted reddening
gives us $(B - V)_{0,g} = 0.83 \pm 0.01$. The magnitude at which $(B - V)_{0} =
1.4$ is found to be $V_{1.4} = 12.422 \pm 0.022$, which gives a value of
$\Delta V_{1.4} = 2.66 \pm 0.02$. The primary source of error in $V_{1.4}$ and
$(B - V)_{0,g}$ is the uncertainty in the reddening, which we take to
be $\pm 0.01$. 

We can also
evaluate the same quantities for other clusters of similar metallicity.
We include NGC 288 (Bergbusch 1993), NGC 362 (Harris 1982), NGC 1261
(Ferraro et al. 1993), NGC 1851 (Walker 1992a), NGC 2808 (Ferraro et
al. 1990), M4 (NGC
6121; Kanatas et al. 1995), and Palomar 5 (C1513+000; Smith et al.
1986), and give the calculated values in Table~6.
Comparison of $\Delta V_{1.4}$ values indicates that the total scatter among
the seven clusters corresponds to about 0.35
dex in [Fe/H] according to the linear relationship in ZW84
(0.2 dex if M4 is left out). By the
$(B-V)_{0,g}$ measure, the total scatter corresponds to a
scatter of about 0.3 dex in [Fe/H]. M5 does appear to have
a metallicity consistent with this group of clusters. We must keep in
mind that the RGB above the level of the HB may be abnormally blue, or
the HB may be abnormally faint, since the $\Delta V_{1.4}$ measurement
seems secure. As ZW84 noted, M3 (Buonanno et al.
(1994) and M5 have
very similar $\Delta V_{1.4}$ values, despite having metallicities that
appear to differ by 0.27 dex by other measures.

We can also compare the upper RGB as measured in $(I,V-I)$ with
those of M15, NGC 6397, M2, NGC 6752, NGC 1851, and 47 Tucanae, as
presented in Da Costa
\& Armandroff (1990). (As a reminder, the metallicities in that study
use the ZW84 zero point.) A disadvantage of using this
method is that it also requires a distance modulus --- we use the
value derived in \S \ref{subd} via subdwarf fitting to the MS, corrected for
the difference between $A_{V}$ and $A_{I}$ (using the reddening law
given by Cardelli, Clayton \& Mathis 1989). (In deciding not to use
a distance modulus derived from fitting the HB, we are breaking with
Da Costa \& Armandroff's distance scale because of the
possibility that M5's HB may be abnormally faint.) A 
comparison is shown in Figure~\ref{fig22}. The positions of the RGBs indicate
that M5 has a metallicity slightly higher than those of  
NGC 6752 ([Fe/H] = --1.54; Armandroff \& Zinn 1988) and M2 ([Fe/H] =
--1.58).

Alternately we can apply a lower value of the reddening (E$(V-I) =
0.03$, corresponding to E$(B-V) = 0.02$) with the corresponding
distance modulus. When this is done, we find that the metallicity
appears to be closer to that of NGC 1851 ([Fe/H] = --1.29; Armandroff
\& Zinn 1988), The agreement of the RGBs in slope is better in this
case. Using the RGB fiducials as reference points, we can constrain
M5's metallicity to the range $-1.5 \gtrsim \mbox{[Fe/H]} \gtrsim -1.3$,
which nicely brackets the ZW84 value of --1.40.

On the basis of these considerations, we find that M5 is
one of the most metal-poor members of this metallicity group (listed in
Table~6) in a relative sense.

\subsubsection{Abundances of $\alpha$-Elements}\label{alph}

M5, like other globular clusters, is observed to have
enhancements in the abundances of the $\alpha$-elements (Pilachowski,
Olszewski \& Odell 1983; Gratton, Quarta \& Ortolani 1986; Sneden et al
1992). The overall enhancement
of $\alpha$-element abundances makes scaled-solar abundance isochrones
appear, to good
accuracy, slightly more metal-rich, or in other words, redder and
fainter (Chieffi, Straniero \& Salaris 1991; Chaboyer, Sarajedini \&
Demarque 1992) This shift is
primarily due to increased envelope H$^{-}$ opacity due to electron
contributions from low
ionization-potential $\alpha$-elements. To first order, it is possible to 
remove the effects of $\alpha$-element enhancement from consideration except
when we must discuss characteristics that
differentiate between [Fe/H] and [$M$/H], the overall metal content
relative to the sun.

Figure~\ref{fig23} shows a comparison of a preliminary
$\alpha$-enhanced isochrone ([$\alpha$/Fe] $= +0.3$;
VandenBerg 1995) with
oxygen-enhanced isochrones (BV92). There are several features to
notice here. First, to very good accuracy, the MS of the
$\alpha$-enhanced isochrone parallels those of the oxygen-enhanced
isochrones. Second, the RGB of the $\alpha$-enhanced isochrone
initially parallels the oxygen-enhanced ones, but becomes slightly
steeper toward the RGB tip. This probably results from the
recombination of electrons with certain of the
$\alpha$-elements at the lower temperatures,
which helps to reduce the H$^{-}$ opacity at the surface. Oxygen does not
contribute to the surface opacity on the RGB due to its relatively
high ionization potential. For surface
temperatures near that of the turnoff, oxygen does
make a contribution fairly near the surface, making the turnoff
redder (Rood 1981, Rood \& Crocker 1993). In the figure, the
$\alpha$-enhanced and oxygen-enhanced isochrones remain parallel
through the turnoff. (For the most metal-poor BV92 
isochrones, the oxygen enhancement continues to increase, meaning that the
oxygen-enhanced isochrones will tend to get redder at the turnoff
relative to constant $\alpha$-enhancement isochrones. This does not
affect the metallicities we will be dealing with.) At the same
time, the increased energy generation from the CNO
cycle actually drives a slight {\it decrease} in luminosity on the subgiant
branch (Rood 1981). {\it As a result, the main shape differences
between the oxygen-enhanced and $\alpha$-enhanced isochrones occur
only in the luminosity of the subgiant branch and the color of the
upper RGB.} Based on these details we will make our analysis using the
complete set of oxygen-enhanced isochrones, keeping in mind the model
differences. 

The effect of an $\alpha$-enhanced composition on the LF can be seen
in Figure~\ref{fig24}. The enhancement has the effect of shifting the SGB
fainter and decreasing the luminosity of the RG bump. With the
exception of the luminosity of the RG bump, and the exact height of the
SGB jump, the $\alpha$-enhanced LF matches an
oxygen-enhanced LF 0.3 dex more metal-rich. The different
heights of the SGB jump indicate a slight difference in the slope of
the SGB in the CMD (see \S \ref{sgblf}). Each of these effects is
consistent with
making the $\alpha$-enhanced models appear much like an
oxygen-enhanced model of higher metallicity.


There is some uncertainty in the absolute amount of $\alpha$-element
enhancement, and this should be kept in mind.
According to Chieffi et al. (1991),
an enhancement in $\alpha$-element abundance by
a factor $f_{\alpha}$ changes the effective metal content according to
the formula
\[ \mbox{[$M$/H]} = \mbox{[Fe/H]} + \log(0.638 f_{\alpha} + 0.362) .\]
For the $\alpha$-enhancement observed in M5 by
Sneden et al. ([$\alpha$/Fe] $\sim +0.2$), a correction of --0.14 is
needed to bring the [$M$/H] value
to an [Fe/H] scale, if we use the relation given by Salaris, Chieffi
\& Straniero (1993). The Sneden et al. result is consistent with an
average enhancement of [$\alpha$/Fe] $= +0.3$ seen for a number of
globular clusters (see Kraft et al. 1995 for references). From here
on, when we consider the ZW84 metallicity scale, we will use
[Fe/H] $= -1.40$ and [$M$/H] $= -1.19$, and when we consider the
Lick-Texas group scale (Sneden et al. 1992), we will use
[Fe/H] $= -1.17$ and [$M$/H] $= -1.03$.

\subsection{Helium Abundance}\label{he}

In the following sections, we re-examine the helium abundance
indicators for M5 and for other clusters of similar metallicity.

\subsubsection{Population Ratios and the $R$ Indicator}\label{rats}

In post-main-sequence phases of evolution, numbers of stars are
directly proportional to the evolutionary timescale, and as a result,
the ratios of populations in different phases reflect the efficiency
of energy generation and mixing processes. The most used ratios are $R =
N_{HB}/N_{RGB}, R^{\prime} = N_{HB}/(N_{RGB} + N_{AGB}), R_{1} =
N_{AGB}/N_{RGB},
R_{2} = N_{AGB}/N_{HB}$, and $R_{HB} = (N_{BHB} - N_{RHB}) / (N_{BHB} +
N_{RR} + N_{RHB})$, where $N_{BHB}$, $N_{RR}$, and $N_{RHB}$ are
the numbers of blue horizontal branch, RR Lyrae variable, and red
horizontal branch stars respectively.
The ratios $R$ and $R^{\prime}$ have been used as an indicator of helium
content $Y$ (Iben 1968, Buzzoni et al. 1983). This primarily reflects the
dependence of the progress of
the hydrogen-burning shell on the hydrogen content of the envelope
material being fed in (and to a lesser extent, the change of helium
core mass at helium flash, which affects the HB luminosity). $R_{1}$
and $R_{2}$ have been used to confirm the existence of
semiconvective zones outside the helium-burning core of HB stars
(Renzini \& Fusi Pecci 1988), as
such zones feed extra helium into the core, prolonging the HB phase. The
values we calculate for $R_{1}$ and $R_{2}$ fall within the range of
values published for clusters over a range of metallicities. This
indicates that HB stars develop semiconvective zones during their
evolution that mix extra helium into the core, prolonging their stay
on the HB. The ratio $R_{HB}$ (Lee 1989) was intended to be a
quantitative indication of HB morphology.

Because the stars in these phases are bright, it
is possible to calculate these population ratios for any sample of stellar
photometry covering the entire HB, and the samples used are
more likely to be complete even into the center of the cluster. We
have chosen to use
the $(I,B-I)$ dataset because of the larger sample, better
seeing, and the superior temperature resolution of $(B - I)$. This
allows us to
use stars over a wider range of radii, as even stars whose
measurements have been somewhat contaminated can be identified with their true
evolutionary phase with high confidence.

Photometric scatter due to crowding effects was significant enough in
the innermost $30^{\prime\prime}$ of the CTIO 4 m data that this
region was excluded from consideration. Known field
stars (Cudworth 1979, Rees 1993) have also been removed from the data.
The results of population counts and ratios are
shown in Table~7. The RR Lyrae sample was composed of those stars
known to be variable (Sawyer Hogg 1973; Kravtsov 1988, 1991; Kadla
et al. 1987), and stars determined to be
variable to good probability (see \S \ref{rr}). The RGB sample is defined
as the stars brighter than the
average bolometric luminosity of the RR Lyrae variables. To define the cutoff
magnitude for our sample, we used the HB models of Dorman (1992b) to
determine the average luminosity and absolute $V$ magnitude for the RR
Lyraes (defined as where $\log T_{eff} = 3.85$), and the isochrones of
BV92 to determine the corresponding
quantities for the RGB. These two theoretical studies form a set of
evolutionary tracks with consistent physics, so that the values derived
should be correct in a differential sense. Using the [Fe/H] $= -1.26$
tracks, we find $\log (L/L_{\odot})_{HB} = 1.645$, $M_{V,HB} = 0.626$,
and $M_{V,RGB} = 0.891$, giving us a bolometric correction of 0.265
mag in $V$. (A similar calculation produces 0.23 mag for M3's
metallicity, in good agreement with the bolometric
correction of 0.22 mag used by Buonanno et al. (1994).) From the $(V,B-V)$
dataset, we determined $V_{HB}$,
and applied the bolometric correction. So, based on this, we used
$B_{RGB} = 16.21$ for the faint end of the RGB population.

To reduce the number of misidentifications among RGB and AGB stars,
we compared with samples taken from uncalibrated data
for the core of M5 from the Canada-France-Hawaii Telescope (CFHT) with the
High-Resolution Camera in April, 1993. The higher resolution
allows us to make nearly unambiguous decisions as to which population
a bright giant star belongs to. In the comparison, we found no
stars misidentified as RGB in the CTIO data set and AGB in
the CFHT dataset, but found 8 stars given as AGB in the CTIO set and RGB in the
CFHT data. This sort of bias is understandable, as the effect of
blending is to increase the measured brightness of a star, so that it
is easier to make AGB-like objects from RGB stars than vice versa. We have
corrected for these misidentified stars in our population ratios, but
the effect is undoubtedly present to some small extent in our data (and
the datasets of other studies), as the
field of view of the CFHT frames was small enough (2\Min2 $\times$
2\Min2) that we were unable to correct the whole core region for
this kind of effect. The CFHT frames are roughly centered on the
cluster center, so that we are getting the region most affected.

We compare our samples to those of BCF81 and Brocato et al. (1995;
hereafter BCR). BCF81's numbers have been corrected for two RR Lyraes
from Sawyer Hogg (1973) that were included in their list: I-104 (V21,
identified as a blue HB star) and IV-96 (V76, identified as a red HB
star). The overall agreement is quite good --- there were some changes
that resulted from improved photometry of several stars. The majority of
the stars that we moved were actually red giants. This partly explains
the large
discrepancy between the numbers of RGB stars. Most of the remaining
difference stems from the larger bolometric correction we used: 0.27
mag instead of the ``average'' bolometric correction for clusters of
0.15 mag (Buzzoni et al. 1983). The larger bolometric correction is
more valid because the theoretical calibrations of the $R$ method
implicitly assume that the RGB cutoff is equal to the mean {\it luminosity}
of the RR Lyraes.
If we use the smaller bolometric correction, our RGB sample of BCF81
stars would be 149. A higher metallicity would result in a larger
bolometric correction. We also include our
sample for the region for which BCR tabulated data. It appears
that BCR are significantly incomplete toward the center of the cluster, since
we find many more RGB, HB, and AGB stars than they do. The surpluses
even show up when we compare with only the CTIO 4 m data set over
the radial range $30^{\prime\prime} < r < 120^{\prime\prime}$. 
This result is surprising since such bright stars should be visible
into the center of the cluster (even though they may be difficult to
identify with the correct phase of evolution). Curiously though, our
sample also shows high values of $R$ and $R_{1}$ over this range of radii
relative to the rest of the cluster. This should indicate that there
are fairly large fluctuations in the sky distribution of the different
populations, and that this kind of fluctuation can change the
calculated population ratios substantially. We did not see signs
of significant radial trends in the population ratios in our sample, however. 

In re-examining the determinations of $R$ for other clusters in the same
metallicity bin, we have found that the vast majority of calculations
have used the ``average'' bolometric correction of 0.15 mag in $V$ in
determining the faint end of the RGB sample. This value 
was used in the original Buzzoni et al. (1983) paper, but
only because uncertainties of various sorts made the correction
unimportant. However, the correction is a function of metallicity, and
it is applied to the faint end of the RGB sample, which means this can
introduce a significant systematic error. In Table~6, we attempt to
correct the situation for the clusters in this metallicity bin. We
include values for NGC 288 (Buonanno et al. 1984), NGC 362 (Harris
1982), NGC 1261 (C0310-554; Ferraro et al. 1993), NGC 1851 (C0512-400;
Walker 1992a), NGC 
2808 (C0911-646; Ferraro et al. 1990), and M4 (NGC 6121 = C1620-264;
Cudworth \& Rees
1990). Once this correction has been made, all of the $R$ values are
consistent with each other to within the errors. They are also low with
respect to that of M3 (Buonanno et al. 1994), and slightly high compared to
that of 47 Tuc (Lee 1977). This could reflect a metal-dependent trend.

The values we derive for $R$ and $R^{\prime}$ are quite a
bit lower than typically quoted values for other clusters (Buzzoni et
al. 1983; Buonanno, Corsi \& Fusi Pecci 1985). However, the helium
abundance derived for M5 in
the Buzzoni et al. study ($Y = 0.20$) was also the lowest found for
their sample of clusters. Using equation 11 of
Buzzoni et al. for our total CTIO sample, we get $Y = 0.18 \pm 0.02$.
For the combined CTIO and CFHT sample, we get $Y = 0.20 \pm 0.02$.
(The error refers only to the statistical error in the population
counts.) Identical results are derived from equation 12 of Buzzoni
et al. for $R^{\prime}$, and no significant change occurs if we use the
calibration of Caputo, Martinez Roger \& Paez (1987) for canonical
assumptions about the mass of the helium cores in HB stars. (The
absolute calibration of helium is somewhat uncertain, as it is unclear
whether non-canonical assumptions, such as enhanced He-core mass for
HB stars, are important.)
For the M5 $R$ value to match the mean value of 1.35 found for
clusters tabulated by Buzzoni et al. (1983), we
would have to have missed 112 HB stars, or counted 83 RGB stars too
many --- requiring a change of over 17\% in the sample for either case
separately. (If we use the smaller bolometric correction, the RGB
sample decreases from 494 to 452 stars, still requiring changes of
approximately 10\% to match the M3 ratio.) As can be seen from our
incompleteness
calculations for the LF, we should be 100\% complete for stars this bright.

Figures~\ref{fig17} and \ref{fig18} should also make it clear that
misidentification is not a possible
explanation. Figure~\ref{fig17} plots all stars that have been
included in the RGB, HB (excluding RR Lyrae stars), and AGB
populations, and that also have calibrated 
magnitudes from the CTIO 4 m data. For the majority of the stars
plotted, the population was assigned based on position in the CMD.
Also included are stars that were measured in both the CTIO 4 m
and CFHT data. The population assignment for these stars was based
on position in the CFHT CMD, while the magnitude and color plotted
correspond to the CTIO 4 m measurements. This is why, for example,
a number of blue HB stars appear in the instability strip, and why
some RGB stars appear closer to the AGB. Misidentification is only
a problem between the RGB and AGB, and for the overwhelming
majority of these stars, photometric scatter or blending effects
do not make any difference. Figure~\ref{fig18} plots the CFHT data (12739
objects total in the frames) for the
cluster core. The sequences in this data are extremely tight, and
misidentification is not a factor. (A
population of blue stragglers can be seen in the figure as well).

To examine the possibility that individual populations have been
affected differently by cluster dynamics, 
we plot the cumulative radial distributions for the RGB,
AGB, and HB stars used in the determination of the population ratios
in Figure~\ref{fig19}. This is the most extensive radial distribution published
to date, as it completely covers $8^{\prime}$ in projected radius.
All three distributions agree extremely well even into the center of
the cluster, although the AGB stars may be slightly less
centrally concentrated compared to the others.
A K-S test indicates that the AGB stars belong to the same radial
distribution as the RGB and HB stars at 66\% and 64\% probabilities
respectively. (For comparison, the RGB and HB stars have an 80\%
probability.) The sample
of AGB stars is large enough that statistical fluctuations are
relatively unimportant.
AGB star samples have in general not been large enough to make
significant comparisons with RGB or HB stars, even in post-core
collapse (PCC) clusters. There is some evidence for deficiencies of RGB
stars compared to HB stars in the PCC cluster M30 (Djorgovski \&
Piotto 1993), but only in the innermost $20^{\prime\prime}$ or so. As M5 has a
King-model surface brightness profile (Trager et al. 1995), it is not
surprising that there are no obvious dynamical effects evident in the
distributions. None of the clusters of similar metallicity with good data
seem to be in a post-core collapse phase, with the
possible exception of NGC 362 (Trager et al. 1995). The
clusters used in these calculations do cover a wide range of stellar
densities though.

\subsubsection{RR Lyrae Stars and the $A$ Indicator}\label{rr}

M5 is one of the most populous globular clusters in terms of RR Lyrae
stars, and yet very few of these stars have been studied extensively
enough to understand M5's contribution to the Sandage period-shift
effect (Sandage 1982). Although our photometry is not extensive enough
to measure
light curves, we did conduct a search for unidentified variables. As
M5 is one of the nearest clusters and has moderate central
density, our spatial resolution
is enough to examine even the core of the cluster.

Coordinates from the list of Sawyer Hogg (1973) were transformed to the
system of our frames, and stars were matched with the list based on
this position, with a check to verify that the magnitudes agreed
roughly. The 87 listed variable stars that fell within the frames
were successfully identified (although apparently two digits in the Y
coordinate for V39 were transposed --- the correct value is $-250.2$).
Two of the 87 variables are not RR Lyraes. In addition, the
stars V102 and V103 that were listed as potential variables in Sawyer
Hogg's list have been confirmed as RR Lyrae variables by Kravtsov
(1992).

For HB stars with $V$ measurements, we determined the ratio of
external magnitude error (determined from
scatter in the measured magnitude values) to internal error
(determined from deviations of the stellar profile from the
point-spread function for the frames). Stars for which $\sigma_{ext} /
\sigma_{int} > 3.5$ were identified as variable. This level was chosen
based on the {\it maximum} background level for red giant stars of
comparable magnitude. From these cuts, we identified 25 RR
Lyrae star candidates --- their coordinates on the system of Sawyer Hogg
are included
in Table~8. Most of these stars had been discovered previously by Kadla
et al. (1987) or Kravtsov (1988, 1991). We have included stars from
these three studies that we have not detected, so as to continue the
numbering of Sawyer Hogg and these studies. Virtually all of these stars are in
the innermost $1^{\prime}$ of 
the cluster --- a region practically unreachable by the early photographic
studies. Judging by the accuracy of the transformation from the Sawyer
Hogg coordinate system to our own for the known variables, we estimate
that these positions should be accurate to better than an arcsecond.

Based on the number of previously identified RR Lyraes that satisfied our
cuts, we can calculate the completeness of this new sample, assuming
that nondetection is due only to stars having unfortunate combinations
of period and phase. Of the 80 RR Lyraes from the Sawyer Hogg list
that had positions that put them in the $V$ frames and magnitudes that
fell in the selected range, 61 were detected by the $\sigma_{ext} /
\sigma_{int}$ cut, giving us a completeness of 76\%. Of the samples of
RR Lyrae candidates in Kadla et al. (1987) and Kravtsov (1988), we
only recover 54\% and 51\% respectively, which may indicate that
some of their candidates are not intrinsically variable. From what
we have been able
to understand of the two papers, not all of the variables have
measured periods. However, we are unable to comment further on the
reliability of their lists.

The helium indicator $A$ (Caputo, Cayrel \& Cayrel de Strobel 1983) is
related to the mass-luminosity relationship for stars inside the
instability strip: 
 \[ A = \log (L/L_{\odot}) - 0.81 \log (M/M_{\odot}) .\]  $A$ has a
relatively small sensitivity to helium abundance ($\partial A / \partial Y
= 1.2$; Caputo et al.
1983), and also requires periods and pulsation
amplitudes for cluster variables, if there are any. We can
calculate values of $A$ for RR Lyrae stars in M5 measured by SCB, and using
equation 16 of Carney, Storm \& Jones (1992)
(a relation between effective temperature, period, B-band
amplitude, and metallicity) and the period relation of van Albada \&
Baker (1971). We restrict ourselves to tabulated RRab variables
(V7, V8, V18, V19, V28, V30 and V32) due to the better determined
effective temperature relation. We find $A = 1.80 \pm 0.02$, where
the error is just the error in the mean. Comparing with values
tabulated in the HB models of Dorman (1992b), we find that this value
is slightly low, but consistent with $Y \approx 0.23$ to within the
errors. We can also calculate values for RRab variables in
NGC 1261 (10 variables; Wehlau \& Demers 1977), NGC 1851 (15 variables;
Wehlau et al. 1978, Wehlau et al. 1982), NGC 2808 (1 variable; Clement
\& Hazen 1989), M4 (15 variables; tabulated in Sandage 1990), M14 (40
variables; Wehlau \& Froelich 1994), M3, and M15
(36 and 25 variables respectively; tabulated in Sandage 1990). The
values are again shown in Table~6. (For M15
we find $1.87 \pm 0.02$.)
Among the clusters tabulated here, there appears to be a trend in $A$ with
[Fe/H], which is not expected. In any case, we have
no evidence from this indicator that the helium abundance in M5 is smaller
than the norm for the globular cluster system.

\subsubsection{The $\Delta$ Indicator}

The indicator $\Delta$ (Caputo, Cayrel \& Cayrel de Strobel 1983) is
defined simply as the magnitude difference between the MS at $(B -
V)_{0} = 0.7$ and the HB at the instability strip. (Caputo et
al. originally defined the HB point to be at the blue edge of the
instability strip. We have chosen to revise this definition to make it
more easily calculable theoretically and observationally. The
various sensitivities of the indicator are unchanged by this revision.) As
an indicator, it has the advantage of being fairly sensitive to the
helium abundance ($\partial \Delta / \partial Y = 5.8$ mag; Caputo et al.
1983), and the disadvantages of having a definite
metallicity dependence ($\partial \Delta / \partial \mbox{[Fe/H]} =
-1$ mag / dex) and of requiring photometry from the HB to well
below the MS turnoff. As found in \S \ref{met},
$V_{HB,be} = 15.079 \pm 0.029$, and from the fiducial line we find
that $V_{0.7} = 20.88 \pm 0.03$. Thus, $\Delta = 5.80 \pm 0.04$. We
can calculate theoretical values of $\Delta$ using a self-consistent set of
isochrones (BV92) and HB models (Dorman 1992b)
having $Y \approx 0.236$ in order to check the value in M5. For
metallicities [$M$/H] = --0.47, --0.65, --0.78, --1.03, --1.26,
--1.48, --1.66, --1.78, --2.03, and --2.26, we find values for
$\Delta$ of 5.17, 5.38, 5.53, 5.76, 5.92, 6.07, 6.14, 6.21, 6.30, and
6.38 respectively. Regardless of the value of [$M$/H] chosen for M5, 
the helium abundance indicated by $\Delta$ is consistent with
the values used in the theoretical models to within the errors. We
also calculate
this quantity for some additional clusters: NGC 288, NGC 1851, 47
Tucanae, M3, and M15 (Durrell \& Harris 1993; $\Delta = 6.29 \pm
0.05$). These values are tabulated in Table~6. With the exception of
NGC 1851, these values agree very well with the theoretical predictions
(for M15, [$M$/H] $\approx -1.89$, for
M3, [$M$/H] $\approx -1.45$, and for 47 Tuc, [$M$/H] $\approx -0.6$
are appropriate). There does not appear to be an obvious trend with
[Fe/H] among the clusters discussed here. 

\subsubsection{Other Indicators}

A relative indication of $Y$ is found in the slope of the blue end of
the HB (Caputo, Natta \& Castellani 1973). The slope of the
blue end of the ZAHB is extremely insensitive to metallicity, as can be
seen in Dorman's (1992b) models. The
slope does vary significantly with the initial helium content given to
the stars, in the sense that helium-deficient clusters should have
more gently sloping (bluer) blue ends, and should have a fainter HB
overall (Sweigart \& Gross 1976; Dorman 1992a). In practice though,
blue HB slope comparisons are untrustworthy due to the scatter in
photometry (making it a difficult task to derive an accurate HB locus) and
color calibrations (typically just a few blue stars are used to calibrate color
terms for globular cluster observations, making color-dependent errors a
real possibility). With these pitfalls in mind, we make several comparisons
in Figure~\ref{fig20}.
The best comparison that can be made is with the ``second-parameter''
cluster NGC 288.  Bergbusch (1993) derives a
helium abundance $Y = 0.232^{+0.020}_{-0.036}$ for
NGC 288 via the $R$-method using a much smaller sample of stars
($N_{RGB} = 56, N_{AGB} = 11$, and $N_{HB} = 77$).
If we shift the CMD using offsets derived from the
relative age technique of VandenBerg, Bolte \& Stetson (1990;
hereafter VBS), we can
look for the effects of different relative $Y$ values. The
slope of the blue end is slightly smaller for M5, although not
significantly so. Additional comparisons can be
made with NGC 1851 (Walker 1992a) and M3
(Buonanno et al. 1994) on the blue side of the HB. NGC 1851 shows a
marginally steeper slope, while M3 shows a shallower slope (although we
must question why the {\it overall} slope of the HB is so much different
from the other clusters examined). This test does not seem to indicate a
significant difference in helium abundance.

From pulsational theory (Deupree 1977), it is expected that the width of
the instability strip should increase with decreased helium content.
Restricting ourselves to the stars with the lowest color errors, we find
that the observed RR Lyrae gap for M5 lies between $(B-V) = 0.24 \pm
0.02$ and $0.47 \pm 0.01$. The
width derived from M3 is virtually identical: $\Delta(B-V)_{RR} = 0.24 \pm
0.03$. The comparison with M3 does not indicate any difference in
helium abundance.
NGC 1851 cannot be used in this comparison as its HB appears to have a
bimodal distribution that does not include the edges of the instability
strip (Walker 1992a), while NGC 288 and NGC 362 do not have HB stars on both
sides of the gap.

We conclude that despite the well-determined fact that M5
has an $R$ value indicating it is deficient in helium, there is no other
evidence in the photometry that would support this. The helium
abundances as derived from the other indicators appear consistent with
determinations from other sources (see Boesgaard \& Steigman 1985 for
a review of older studies). For example, Campbell (1992) finds (He/H)
$= 0.0759 \pm 0.014$ (corresponding approximately to $Y_{P} = 0.233
\pm 0.003$) from a sample of H II galaxies, while Olive \& Steigman
(1995) derive $Y_{P} = 0.232 \pm 0.003$ from helium abundances in metal-poor
extragalactic H II regions, extrapolated to zero metallicity. At the
same time, observations of D and $^{3}$He abundances provide a lower
limit $Y_{P} \geq 0.238$ (Walker et al. 1991). As mentioned
earlier, an unexpected metal-dependent trend in the $R$ values may be
present in the limited sample examined here. An examination of
BV92 isochrones and Dorman (1992b) HB tracks
shows that the metallicity dependence of $R$ is negligible as
Buzzoni et al. (1983) indicate. So, any metallicity dependence is
indirect at best. These possibilities should be examined using a
larger sample of clusters. However, the value of $R$
as a helium indicator is limited by the total stellar populations in
individual clusters, and little improvement can be expected beyond
determinations like the one for M5 presented in this paper. The
indicator $\Delta$ is likely to provide a better handle on helium
abundance in the future.

\subsection{The Distance Modulus}\label{6.2.}

An accurately determined distance modulus for the cluster, combined
with a known reddening value, allows the
fiducial line to be placed in a diagram of absolute magnitude versus
color, and compared directly with theoretical isochrones. In
principle, the age of the cluster could be determined simply from this
comparison. However, any distance determination requires the
assumption of a value for the metallicity.

\subsubsection{Subdwarf Fitting}\label{subd}

Subdwarfs having known metallicities and well-measured trigonometric
parallaxes provide a calibration of the brightness of the main
sequence fiducial of the cluster. Our sample of 23 potentially-usable
subdwarfs (along with 5 subgiant stars) is provided
in Table~9. The first 9 subdwarfs (which we call the ``classical''
subdwarfs) are taken from Table II of Laird, Carney \& Latham
(1988). Due to M5's high
metallicity, we opted to include the star HD23439A ([Fe/H]
= --1.02). Each of these stars has relative parallax errors of $\sigma_{\pi}
/ \pi \leq 0.12$. We have supplemented this list with other subdwarfs
identified from the literature as having
relatively small errors in their parallaxes ($\sigma_{\pi} / \pi \leq
0.30$). These stars are identified as halo population dwarfs by
their metallicities ([Fe/H] $\gtrsim -1.0$). Five subgiants that pass
the cut on relative parallax error are also included.

The $V$ magnitudes and $(B-V)$ colors of these stars were taken from the
Hipparcos Input Catalogue (HIC; Turon et al. 1993). The
values for the classical subdwarfs agree with those tabulated in
Laird et al. In any case, the accuracy of these values is certainly not the
major contribution to the error in this method --- parallax error is.
The parallaxes were taken from the General Catalogue of Trigonometric
Parallaxes (van Altena, Truen-liang Lee \& Hoffleit 1991). The
parallaxes of 15 stars used the values given in Dahn (1994), as
they often represented significant improvements, as judged by the agreement
with theoretical isochrones.

The absolute magnitudes were calculated
using Lutz-Kelker corrections (Lutz \& Kelker 1973) derived by
the method of Hanson (1979). These corrections are intended to rectify
a systematic
bias in samples selected by proper motion or trigonometric parallax ---
stars with parallaxes measured too high are more likely to be included
in the sample. The original Lutz-Kelker corrections were
derived with the assumption that the sample stars have a uniform space
density. This assumption implies particular parallax and proper motion
distributions (a probability distribution $P(\pi) \propto \pi^{-4}$
and a number distribution $N(\mu) \propto
\mu^{-3}$, where $\mu$ is proper motion). As a different spatial
distribution would change the values
of the corrections, following Hanson, we examined the proper motion
distribution of our subdwarf sample using values from Luyten (1976)
and found $N(\mu) \propto
\mu^{-1.2}$. This result implies that proper motion and luminosity (or
Malmquist)
selection biases are present in the sample in addition to the parallax
selection bias. These effects act to reduce the Lutz-Kelker
corrections below the constant-density relation, since proper motion
and luminosity selection biases tend to discriminate against distant
stars --- an effect opposite to the parallax selection bias. Because
subdwarfs are selected largely on the basis of high proper motions,
this bias is probably the more important. As a result of this
finding, we have used Hanson's relation for $n = 2$ (where $P(\pi)
\propto \pi^{-n}$ and $N(\mu) \propto \mu^{-(n-1)}$) to calculate the
magnitude corrections. For
the classical subdwarfs, these corrections are very small ($-0.08 \leq
\Delta M_{LK} \leq 0$). 

The metallicities for the subdwarfs
were taken from Carney et al. (1994) when available, from Laird et
al. (1988), Ryan \& Norris (1991), and from literature values as
compiled in Cayrel de Strobel et al. (1992) in other cases. The Ryan \& Norris
metallicity data were calibrated using values measured by Laird et
al., so we can be reasonably assured that any systematic differences
are small. We did change the metallicity value for only one star (HD
201891) that had a value in the Carney et al. survey, and only because
all other measurements indicated a metallicity that was higher by 0.2 -
0.4 dex. For the subgiant stars (excluding HD 140283), we take the
metallicities from Pilachowski, Sneden \& Booth (1993) with a
correction of --0.35 to account for measured zero-point differences.

There is a large body of evidence (see Lambert 1989) that indicates the
$\alpha$-element
abundances of the subdwarfs are enhanced in a manner similar to those
of the globular clusters. In order to check how well the colors of the
subdwarfs are predicted, we corrected the metallicities of the
subdwarfs for an enhancement [$\alpha$/Fe] $\sim +0.3$, which
corresponds to [$M$/H] = [Fe/H] + 0.21.
Using a polynomial interpolation between the oxygen-enhanced 14 Gyr
isochrones of BV92, we calculated the colors
corresponding to [$M$/H] for the subdwarf 
and to the value of [$M$/H] for M5 for the absolute $V$
magnitude of each subdwarf. The corrected colors
and deviations of the measured colors from the relevant isochrone
($\delta_{B-V}$) are
presented in Table~9, while the best fit of the M5 fiducial to these
stars is shown in Figure~\ref{fig25}. The agreement of the theoretical
with the observed colors is quite good for the classical subdwarfs:
$\delta(B-V) = 0.010\pm0.015$.
About half of the other candidates have error bars that overlap with
the predicted position. The fit for the subgiant HD140283 is
reasonably good considering the controversy over its distance. The
subgiant HD211998 also matches the predictions quite well.
Although age differences among the stars could be problematic, these
stars are potentially valuable in assessing whether the
isochrones can predict giant branch colors in addition to those for
the main sequence. With the release of Hipparcos parallax data in the
near future, a set of well-measured subgiants could potentially provide a
useful check of the variation of the mixing-length
parameter $\alpha$ with metallicity.

There appears to be a systematic color offset between
a number of the subdwarfs and the theoretical predictions, as shown in
Figure~\ref{fig26}. The offset is approximately +0.08, and is in the
sense that the
subdwarfs are redder than their predicted positions. This difference
cannot be explained by reddening (only one of the stars for which we have
reddenings has $E(B-V)$ as large as 0.02). A systematic shift in the
metallicity scale by about 0.6 dex would be required to explain the
difference, which is unlikely given that the quoted errors
in the metallicities are approximately 0.2 dex. Small errors in the
parallax could cause relatively large differences in the derived color
for stars near the turnoff.
This would not explain some of the fainter subdwarfs though. Several
of these deviant subdwarfs have among the best measured parallaxes, which seems
to rule out overestimation of the Lutz-Kelker corrections. The shape
of the $\alpha$-enhanced isochrones in the
region of the turnoff is probably different from the oxygen-enhanced
isochrones we used, but this would only affect subdwarfs near the turnoff.
Potential age differences between the subdwarfs and the
isochrones are very unlikely to be large enough to explain the
discrepancy as main sequence stars more than half a magnitude below the
turnoff do not evolve enough in color
(Figure~15 in BV92). In addition, comparisons of field subdwarf colors
with turnoff colors of globular clusters indicate that the field
subdwarfs have ages approximately equal to those of the halo globular clusters
(Figure~5 in Gilmore, Kuijken \& Wyse 1989; Figures~3-5 in Carney,
Latham \& Laird 1989). Some of the stars might be unresolved binaries,
which could explain many of the deviations. As a result
of this unresolved problem, we decided to use only the classical subdwarfs (BD
$+66^{\circ}268$, HD 23439A, HD 25329, HD 64090, HD 103095, HD 134439,
HD 134440, HD 194598, and HD 201891) in
our determination of the distance modulus, as they agreed fairly well with
the theoretical isochrones without any additional correction.

Given the metallicity of the subdwarf and M5, we computed a color
shift for each subdwarf. The fiducial of M5 was then
shifted in magnitude to match each subdwarf individually. The
distance modulus values derived for each subdwarf in the sample were
then combined in an average weighted by the squares of the error
estimates for the
absolute magnitudes. This error estimate includes the error resulting
from the uncertainty in the parallax, the reddening, and in the
metallicity (taken to be 0.2 dex for each subdwarf). The final error
estimate is simply the error in the weighted mean. For [Fe/H] =
--1.40, the derived apparent
distance moduli for each of the classical subdwarfs are:
BD$+66^{\circ}268$, $14.49\pm0.27$; HD 23439A, $14.45\pm0.21$; HD 25329,
$14.59\pm0.56$; HD 64090, $14.43\pm0.28$; HD 103095, $14.41\pm0.10$;
HD 134439, $14.37\pm0.18$; HD 134440, $14.36\pm0.19$; HD 194598,
$14.52\pm0.43$; and HD 201891, $14.38\pm0.42$.

With these subdwarfs, we derive an apparent distance modulus $(m -
M)_{V} = 14.41\pm0.07$, assuming M5's metal content is [$M$/H] $\sim
-1.19$ ([Fe/H] $\sim -1.40$). The largest uncertainty is due to
potential errors in the metallicity. If we redo the analysis for
[$M$/H] = --1.03 (the Lick-Texas value), --1.26, and --1.48, we derive
apparent moduli $(m - M)_{V} = 14.50 \pm 0.07$, $14.41 \pm 0.06$, and
$14.32 \pm 0.06$ respectively.  If we do not apply Lutz-Kelker
corrections to this subdwarf sample, we derive an apparent modulus of
$14.39\pm0.07$ for [$M$/H] $= -1.19$. If the reddening is reduced to
E$(B-V) = 0.02$, the apparent modulus changes to 14.40. In any case,
the apparent distance modulus is somewhat larger than that derived by
RF ($14.30\pm0.20$), primarily as a result of the systematic color
offset between our photometry and theirs on the MS.  If we correct for
absorption, the true distance modulus is then $(m - M)_{0} = 14.32 \pm
0.07$.  This distance modulus agrees well with Baade-Wesselink
distances to two RR Lyraes: $(m - M)_{0} = 14.37 \pm 0.13$ (Storm,
Carney \& Latham 1994).

\subsubsection{SGB Fitting}

Because the subgiant branch ``jump'' is a very sharp feature in
$B$-band magnitude, a determination of the distance modulus from a
theoretical LF fit (shifting the theoretical LF to match the magnitude
position of the feature) provides a consistency check on the subdwarf
distance --- the estimated distance modulus increases with
increasing metallicity for subdwarfs, and decreases with increasing
metallicity for a SGB fit. The magnitude position has a moderate 
sensitivity to age (0.07 mag / Gyr at this
metallicity in $B$-band; see Figure~\ref{fig31}) and a dependence on
details of the chemical composition (see \ref{alph}), and so should not be considered a primary distance
indicator.  Figure~\ref{fig27} shows the results of this fit for
models with an age of 14 Gyr. For the metal content of $-1.19$, we
would derive an apparent modulus of $(m - M)_{B} \approx 14.4$, which is
slightly lower than the subdwarf value. A smaller age (by about 3 Gyr)
would be needed to bring the results for [$M$/H] $= -1.03$ into
consistency.

\subsubsection{HB Fitting and RR Lyrae $M_{V}$ -- [Fe/H]
Relations}\label{hbfit}

We may attempt to fit theoretical HB models to the observed HB
sequence as another indication of the distance modulus. The best fits
using models from Dorman (1992b) are shown in
Figure~\ref{fig28}. Fitting by eye with the [Fe/H] = --1.26 models, we
get an apparent distance modulus of $14.52\pm0.05$. A difference of metallicity
of 0.2 dex only changes the distance modulus fit by about 0.05
mag. This value is in good agreement with the distance modulus used by
Da Costa \& Armandroff (1990). The overall fit is not as satisfactory
--- the slope of the blue end of the HB is not matched. However, this
may be a result of our color calibration. (The red end is best fit by
the [Fe/H] $= -1.48$ tracks.) The disagreement between the two
distance moduli we have derived cannot be explained by the color
calibration, however. If the difference between the subdwarf and HB
fitting values for the distance modulus is real, this would mostly
explain the earlier disagreement of the value of $\Delta V_{1.4}$ for
M5 with those of other clusters of the same metallicity. Alternately,
the distance modulus from HB fitting would match the value from
subdwarf fitting at a metallicity between the values given by ZW84 and
the Lick-Texas group.

The linear dependence of $M_{V}$ on [Fe/H] for RR Lyraes is also an
important means of determining the distance modulus for distant old
populations via photometric means. The slope and zero-point of the
relation still remain in dispute, however.  Recent analyses seem to
indicate that a low value for the slope ($\partial M_{V}(\mbox{RR}) /
\partial \mbox{[Fe/H]} \approx 0.15$) is favored over a value that
would explain the Sandage period shift effect (0.39; Sandage \&
Cacciari 1990). The zero-point is more in doubt.  Some of recent
determinations of this relation are: Carney et al.  (1992) from a
reevaluation of earlier studies
\[ <M_{V}(\mbox{RR})> = (0.15 \pm 0.01) \mbox{[Fe/H]} + (1.01 \pm 0.08);\]
from field RR Lyrae stars, Jones et al. (1992)
\[ <M_{V}(\mbox{RR})> = (0.16 \pm 0.03) \mbox{[Fe/H]} + (1.02 \pm 0.03)\]
and Skillen et al. (1993)
\[ <M_{V}(\mbox{RR})> = (0.21 \pm 0.05) \mbox{[Fe/H]} + (1.04 \pm 0.10);\]
Lee (1990) using theoretical models for $Y = 0.23$
\[ <M_{V}(\mbox{RR})> = 0.17 \mbox{[Fe/H]} + 0.79;\]
and Walker (1992b) using the Carney et al. slope and the distance to
Large Magellanic Cloud (LMC) cluster variables
\[ <M_{V}(\mbox{RR})> = (0.15 \pm 0.01) \mbox{[Fe/H]} + (0.73 \pm 0.10).\]
If the distance modulus for the LMC is revised from 18.50 to $18.37
\pm 0.04$ as suggested by a reanalysis of data from the light rings of
SN1987A (Gould 1995), this zero-point would become $0.86\pm0.01$.

Given these uncertainties, we instead calculate $M_{V}$(HB) for M5
using distance moduli calculated from subdwarf fitting, and compare
with values for various $M_{V}$(RR) relations. Using [$M$/H] $= -1.19$
and $V_{HB} = 15.092 \pm 0.02$, we find that $M_{V}(\mbox{HB}) =
0.68$, which should correspond to the mean magnitude level of the RR
Lyraes. The Carney et al., Jones et al. and Skillen et al. relations
all predict HB magnitudes that are fainter than observed by 0.10 -
0.15 mag, predicting an apparent distance modulus $(m - M)_{V} \approx
14.3$. The Lee theoretical models and the Walker relation predict an
HB that is too bright by about 0.10 mag and 0.13 mag
respectively. (The error in these distance moduli is approximately
0.12 mag.) For this metal content, the value of $M_{V}$(HB) does not
favor either the larger or smaller value for the zero-point.

If we use [$M$/H] $= -1.03$, $M_{V}{\mbox{HB}} = 0.54$. In this case,
Carney et al., Jones et al. and Skillen et al. relations all predict
HB magnitudes that are too faint by about 0.3 mag. In contrast, the
Lee theoretical models are too faint by only 0.07 mag, and the Walker
relation is only faint by 0.03 mag.

\subsubsection{AGB Fitting}

The theoretical HB evolutionary tracks (Dorman 1992b) also allow us to
examine the constraints that the asymptotic giant branch
provides. None of the sets of tracks adequately reproduce all of the
features of the lower AGB. The [Fe/H] $= -1.03$ tracks are marginally
red enough to explain the color of the lower AGB, but the other two
sets are too blue. All of the sets seem to overestimate the luminosity
of the AGB clump as well. Dorman's models would imply an apparent distance
modulus of approximately 14.63.  The position of this feature should
correspond to the range of magnitudes in which the evolutionary track
turns fainter for a time.  (The cause is the envelope adjustment to
the recently-formed helium-burning shell.) The helium-burning models
of Castellani, Chieffi \& Pulone (1991) (their Figure 9) seem to match
the M5 data from BCF81 for both the color and level of the AGB clump
for an apparent modulus of 14.45 and a metallicity [Fe/H] =
--1.13. This metallicity is consistent with M5's metal content if
$\alpha$-element enhancement is taken into account.

\subsection{The Luminosity Functions}\label{lfdisc}

Figures~\ref{fig13} and \ref{fig14} plot the $B$- and $I$-band LFs
along with theoretical LFs derived from BV92 LFs. The theoretical LFs
use distance moduli derived from subdwarf fitting to the main sequence
(see \S \ref{subd}). The theoretical LFs were also degraded to the
finite resolution of the bins in the observed LF.

\subsubsection{The Subgiant Branch}\label{sgblf}

In several metal-poor clusters, so-called ``WIMP bumps'' (or subgiant
excesses) have been observed near this location in the LF. In $B$
band, the peak ($B \approx 18.5$) and dip ($B \approx 18.7$) features
seen in the SGB region of the observed LF in Figure~\ref{fig15} (about
0.4 mag brighter than the turnoff) are affected by the presence of
significant numbers of blends in the photometry. However, we do not
believe that blends could be responsible for the unexplained subgiant
excesses. Blends do not seem to be capable of creating bumps in the
LF, but can only reduce the number of counts in the bins corresponding
to the most horizontal portion of the SGB in the CMD (in the
``peak''). Despite the fact that the blend sequence crosses the the
fiducial sequence of the cluster at approximately this position, no
excess of stars appears there in the LF when crowded regions of the
cluster are included in the sample.

Because the SGB is closest to horizontal in the color-magnitude
diagram when $B$-band is used, any potential subgiant excess would be
confined to the smallest number of magnitude bins, and would thus be
viewed at its greatest significance. We find only one bin exceeding
the theoretical models, with a significance of only about $1
\sigma$. In $I$ band, the SGB is much less horizontal in the CMD, so
we are able to examine the branch in finer detail. The observed counts
follow the theoretical predictions very well, and there is no hint of
a multiple bin enhancement to the counts that would be present if some
physical process was slowing the movement of stars across the
CMD. There also does not appear to be evidence for any excess (or of
significant blending effects) in the $V$-band LF of NGC 288 (Bergbusch
1993), a cluster with metallicity similar to that of M5, but with a
much smaller central density. We conclude that there is no subgiant
excess in the LF for M5, and that this extends to other clusters of
similar metallicity.

The subgiant branch is also the portion of the LF most affected by the
specifics of the chemical composition. The helium abundance exerts a
great effect on the SGB (Simoda 1972), while leaving the relative
numbers of stars on the RGB and MS mostly unchanged (Ratcliff 1987). A
decrease in helium abundance causes the slope of the SGB in the CMD to
become shallower. The SGB fits into fewer bins, enhancing the peak in
the LF. Thus, the peak we see (mimicked by an increase in [Fe/H] in
Figure~\ref{fig13}) may be an indication that the helium abundance in
M5 is low.

\subsubsection{The MS-RGB Discrepancy}\label{5.2.2.}

The discrepancy between the relative numbers of stars observed on the
MS and RGB and the predicted numbers for metal-poor clusters is
notably unresolved. The absolute number of stars on the MS in the LF
is determined by the total number of stars in the cluster and the mass
function from which they are taken. The number of RGB stars is
primarily a function of the evolutionary timescale for those
stars. So, if the discrepancy observed in metal-poor clusters like M92
(Stetson 1991) and M30 (Bolte 1994) is due to errors in
theoretically-determined RG evolutionary timescales, there is a
potential impact on spectral synthesis models, as RGB stars provide a
majority of the integrated light in old stellar populations in
$V$-band and at longer wavelengths.

In the case of M5 though, there does not appear to be a significant
discrepancy. For the $B$-band LF, the agreement with theory is quite
good all of the way up the RGB. The MS and SGB for the $I$-band LF
also match quite well, but the slope of the RGB portion may deviate
from the theoretical prediction. Without the additional evidence
provided by the upper RGB, this deviation might not have been
noticed. The slope difference shows up in the raw data for the
cluster, and it is not due in any way to the procedures used in
calculating the LF. As our photometric calibration showed, we do not
appear to have any magnitude-dependent trends in our photometry in
$I$-band this high on the RGB (see \S \ref{cal}). Possibly this
results from problems in the conversion from bolometric luminosity to
$I$-band magnitude in the models.  In the majority of the fits using
``reasonable'' ages and metallicities though, the MS and RGB are
matched simultaneously. This match is unaffected by our use of
rescaled counts for RGB stars.

The $V$-band LF of the cluster NGC 288 (Bergbusch 1993) also shows
good agreement between the relative levels of the MS and the RGB.
Taken together with the evidence for metal-poor clusters like M92 and
M30, this may be an indication of a metal-dependent trend. A few input
parameters can change the relative levels (Ratcliff 1987). The initial
mass function (IMF) exponent that is used in the theoretical LF can
change the normalization of the MS, but it also changes the shape of
the MS portion of the LF. We have chosen values of the exponent that
match this shape best, and when this is done we do not see a
significant discrepancy.  There is no evidence that the slope of the
MS for the metal-poor clusters is poorly matched.

Helium abundance and metallicity have effects on the relative levels,
but change the morphology of the SGB bump at the same time. Increased
metallicity increases the size of the difference between the MS and
RGB levels, and steepens the ``jump'' at the SGB. Decreased helium
abundance has similar effects.

\subsubsection{The RGB Bump}\label{5.2.3.}

The position of the bump in luminosity directly relates to the point
in a star's evolution at which the hydrogen-burning shell source
passes through the former base of the convection zone (Thomas 1967,
Iben 1968). The increase in the hydrogen content of the material being
fed into the burning shell causes the star to readjust its structure
somewhat, creating a pause in the evolution.  For a sample of 13
clusters, Fusi Pecci et al. (1990) found that the observed bumps were
fainter than predicted by approximately 0.4 magnitudes. This fact does
hold some interest --- it tells us that either the hydrogen burning
shell has moved outward more quickly than expected, or that convection
penetrated more deeply than expected at the base of the RGB. The
former is unlikely to be true, as we have already seen that the MS-RGB
agreement gives no indication that the red giants are evolving
differently than predicted. Thus, the discrepancy between the observed
and predicted positions of the bump may reflect the effects that
non-standard (but physically motivated) processes like convective
overshooting have on the hydrogen profile.

From 107 stars in the $BV$ sample judged to be in the bump, we derive
$V_{bump} = 14.964 \pm 0.007$ ($\Delta V_{bump}^{HB} \equiv V_{bump} -
V_{HB} = -0.13 \pm 0.02$), and $B_{bump} = 15.833 \pm 0.007$.
Figure~\ref{fig16} shows the $B$-band LF in expanded detail near the
observed RG bump.  If we were to simply use [Fe/H] $= -1.40$ for M5
(ZW84), we would find that the bump is fainter than predicted by
approximately 0.3 mag in B (0.25 mag in V), using the subdwarf
distance modulus. A fainter-than-predicted RGB bump can potentially be
explained by a higher cluster age, a higher metal content, or a lower
helium abundance than assumed. From BV92 and Fusi Pecci et al. 1990,
we can estimate some partial derivatives for the approximate
parameters of M5 (age 14 Gyr, [Fe/H] = --1.15, and helium abundance Y =
0.23):
\[ \left( \frac{\partial M_{V}^{bump}}{\partial t_{age}} \right) \approx
0.05 \; \mbox{mag / Gyr} \]
\[ \left( \frac{\partial M_{V}^{bump}}{\partial
[\mbox{Fe}/\mbox{H}]} \right) \approx 0.9 \; \mbox{mag / dex} \]
\[ \left( \frac{\partial M_{V}^{bump}}{\partial Y} \right) \approx
-5 \; \mbox{mag} .\] Typically quoted errors for globular cluster ages
($\pm 3$ Gyr) can only explain about 60\% of the RGB bump
discrepancy here, given a systematic favorable offset. An error in
distance modulus could also be invoked to explain part of the
difference in some previous studies. However, this explanation is
unlikely in this case, since we can judge the position of the RGB bump
relative to the MS turnoff. We find that the difference in magnitude
between bump and turnoff ($\Delta V_{TO}^{bump} = 3.61 \pm 0.05$) is
also smaller than theory predicts ($\Delta V_{TO}^{bump} = 3.88$ for
[Fe/H] $= -1.40$ and age 14 Gyr), independent of distance
modulus. ($\Delta V_{TO}^{bump}$ also has only a small age dependence
of about 0.02 mag per Gyr, so that a large error in our age assumption
would be still be necessary.)

A helium deficiency on the order of 0.04 could explain most of the
disagreement. However, the derivative above fails to take into account
the increased turnoff mass implied by a decreased helium content for a
fixed age (Alongi et al. 1991). An increased turnoff mass causes the
bump to become brighter, which means that less than half of the
disagreement could be explained by this kind of helium deficiency. An
error in the zero-point of the metallicity scale as discussed earlier
could help to remove part of the difference. The Sneden et al. (1992)
value $<$[Fe/H]$> = -1.17 \pm 0.01$ could just be able to account for
all of the difference ($\Delta V_{TO}^{bump} \approx 3.65$).

However, the effects of $\alpha$-element enhancement certainly must be
accounted for. These elements are not as important to the energy
generation for the stars as they are to the envelope opacity. Thus,
the LF would presumably be unaffected by $\alpha$-enhancement {\it
except} for the SGB (as discussed earlier) and the position of the RGB
bump relative to the turnoff of the cluster. Using the formula given
by Chieffi et al. (1991) for determining the enhancement of metal
content $\alpha$-elements, we find that the movement of the bump goes
like \[ \left( \frac{\partial M_{V}^{bump}}{\partial
[\alpha/\mbox{Fe}]} \right) \approx 0.6 \; \mbox{mag / dex} .\] An
enhancement of $+0.3$ dex in the $\alpha$-element abundances would
remove most or all of the discrepancy. For M5, the discrepancy is
reduced to approximately 0.06 mag for $M_{V}$ and $\Delta
V_{bump}^{HB}$ if we use the ZW84 metallicity zero-point ([$M$/H] $=
-1.19$). An additional adjustment of the metallicity zero-point could
make the agreement perfect, or cause the theoretical predictions to
become fainter than the observed bump.

The theoretical prediction for $\Delta V_{TO}^{bump}$ for the metal
content of M5 is then reduced to 3.75. This measure of the bump's
position may be useful, as theoretical predictions can be made without
needing to make assumptions about the HB models. By this comparison,
we see that the bump discrepancy may not be completely resolved ---
relative to the turnoff, the observed bump is still about 0.14 mag too
faint. This may be another indication that M5's HB is abnormally faint
by about 0.1 mag (which would artificially improve the agreement of
$\Delta V_{bump}^{HB}$ with theory), or that the zero-point of the
metallicity scale needs to be closer to the Lick-Texas group
value. However, the magnitude level of the turnoff is also notoriously
difficult to measure, so that part of the discrepancy could be due to
mismeasurement.

Regardless of the case with M5 however, observed $\alpha$-element
enhancements do improve the agreement of the magnitude of the
RG bump with theory for the globular cluster system as a whole. In so
doing, this decreases the need for non-standard mixing processes like
convective overshooting.

\subsection{Age Indicators}\label{6.3.}

Age determination for M5 is of particular interest because M5's
horizontal branch morphology is intermediate between those of the
``second parameter'' pair  NGC 288 and NGC 362 ($R_{HB} = 0.35$ for
M5, versus 0.95 for NGC 288 and --0.87 for NGC 362; Lee 1989). In the
scenario of Lee, Demarque \& Zinn (1994) (with a metallicity [Fe/H] $=
-1.40$ from ZW84), M5 would have an age approximately 1 Gyr older than
the constant age line in their [Fe/H] versus HB Type diagram. The
exact value of such an age offset is highly sensitive to the input
parameters (such as mass loss prescription and chemical composition)
used in the calculations (Catelan \& de Freitas Pacheco
1993). For example, if the metal abundance of M5 is that given by the
Lick-Texas group, the age difference would increase to about 3 Gyr.

In the following analysis, we have chosen not to use the indicator
$\Delta (B - V)$ (Sarajedini \& Demarque 1990; VBS) because the
fiducial lines of the other clusters in this metallicity group are not
determined to high enough accuracy, and color term calibration is not
carried out in a uniform manner for all of the studies.

\subsubsection{$\Delta V_{TO}^{HB}$}\label{delv}

The magnitude difference between the turnoff ($V_{TO} = 18.57\pm0.05$
for M5) and the HB is a widely-used indicator of relative age. From
our photometry, $\Delta V_{TO}^{HB} = 3.47 \pm 0.06$, which agrees
quite well with the value of $3.52 \pm 0.09$ derived for M3 by
Buonanno et al. (1994), and the average of $3.55 \pm 0.09$ for a
sample of 19 clusters examined by Buonanno, Corsi \& Fusi Pecci
(1989). For the other clusters in Table~6, we have chosen to use
studies with the best statistics on both the MS and HB, so that the
quantity is determined as well as possible. We used the following
studies toward this end: NGC 288, Bolte (1992), Bergbusch (1993); NGC
362, Harris (1982), Bolte (1994); NGC 1261, Ferraro et al. (1993); NGC
1851, Walker (1992a); NGC 2808, Alcaino et al. (1990); M4 (NGC 6121),
Kanatas et al. (1995); and Palomar 5, Smith et al. (1986). (Our value
for NGC 2808 is smaller than that stated by Alcaino et al. because
their photometry supports a larger value for $V_{HB}$ than the value
they assumed.)

Because of the color extent of the HB for M5, we can use this dataset
as a ``bridge'' between clusters with widely varying morphology,
enabling us to extend the use of the indicator $\Delta V_{TO}^{HB}$.
We can calculate $\Delta V_{TO}^{HB}$ using another point on the HB
for clusters that do not populate the vicinity of the RR Lyrae gap,
and then do the same for M5 to make a relative age comparison
possible.  If we use the red end of NGC 288's HB to define the
magnitude level of the HB, we derive $\Delta V_{TO}^{HB} =
3.47\pm0.07$ for M5, and $\Delta V_{TO}^{HB} = 3.73$ for Bolte's and
Bergbusch's NGC 288 photometry (approximately $\pm0.09$ for both). For
the color of the blue end of NGC 362's HB, we derive $\Delta
V_{TO}^{HB} = 3.46\pm0.07$ for M5 and $\Delta V_{TO}^{HB} =
3.41\pm0.12$ using the bright star photometry of Harris (1982) and the
smoothed fiducial of VBS for NGC 362. In light of the kind of
zero-point differences between photographic and CCD photometry in
general, the $\Delta V_{TO}^{HB}$ value derived for NGC 362 should be
used with caution. The data for the other clusters, however, are definite
improvements over the same quantities as derived by Buonanno et
al. (1989) simply because the photometry for the MS and HB of each
cluster now comes from a single study.

From the values of $\Delta V_{TO}^{HB}$ alone, there does appear to be
a significant age difference between NGC 288 and M5 (although see the
following discussion). An age difference between NGC 288 and NGC 362
is slightly less secure due to the possibility of large zero-point
differences between the bright and faint samples used for NGC
362. However, a large and systematic error ($\sim +0.1$ mag) in the
zero points of the bright-star and MS photometry studies of NGC 362
would be needed to simply throw doubt on the age ordering. Unmeasured
metallicity differences almost certainly do not play a role, since the
effect on $\Delta V_{TO}^{HB}$ is quite minor. Even if the most
unfavorable and unrealistic $M_{V}^{HB}$ -- [Fe/H] relation is assumed
(constant $M_{V}$; Buonanno et al. 1989), and M5 in reality has a
metallicity 0.3 dex less than that of NGC 288, this only explains
approximately 0.1 mag of the difference in $\Delta V_{TO}^{HB}$. For a
realistic $M_{V}^{HB}$ -- [Fe/H] relation, only 0.05 mag could be
explained by such a metallicity error.

Another influence that must be considered is helium abundance
variations, as an increase in MS helium abundance results in an
increase in the luminosity of the HB, while the effect on the turnoff
is comparatively small ($\partial V_{TO} / \partial Y = -1.4$ mag from
VandenBerg \& Bell isochrones). Our earlier consideration of the
helium abundances now becomes important. From the values of $R$ and
$\Delta$, we find that there is no evidence for differences in helium
abundance greater than 0.02 among M5, NGC 288, and NGC 362. (It is
particularly interesting to note that the values of $\Delta$ for these
three clusters agree much more accurately than the values of $\Delta
V_{TO}^{HB}$. The only real difference between the definitions of the
two indicators is the position of the fainter reference point.) We
estimate that $\Delta V_{TO}^{HB}$ should increase by approximately
0.04 mag for a 0.01 increase in $Y$. So, even if M5 has a helium mass
fraction that is 0.02 lower than NGC 288, we can only explain about
0.08 mag of the difference. In combination with a systematic
unfavorable metallicity error, we can at most explain about half of
the difference.

The evolutionary status of the bluest stars must also be considered,
as the reddest HB stars in clusters with blue HB morphologies are
likely to be evolving toward the AGB at higher luminosities than the
ZAHB (Rood \& Crocker 1993). In the extreme case that all of the HB
stars with $(B-V)_{0} < 0.01$ in NGC 288 have evolved to their
locations from bluer on the HB, conventional evolution tracks predict
the stars would be between 0.1 and 0.4 magnitudes brighter than the
ZAHB, and $\Delta V_{TO}^{HB}$ would be larger by this amount even if
NGC 288 was coeval with the other clusters. It is certainly true that
the red portion of NGC 288's HB does not follow the morphology of M5's
HB, while the blue ends do seem to match each other (along with that
of NGC 1851; see Figure~\ref{fig20}). In addition a gap appears in the
distribution of stars in Bergbusch's study at $(B-V)_{0} \approx
-0.01$ where the two HBs diverge.

However, a coherent well-populated structure is not to be expected
from the evolutionary timescales in theoretical models. Given the 38
stars observed in most-populous portion of NGC 288's HB ($(B-V)
\approx -0.04$; Bergbusch 1993), Dorman's models predict that at most
two stars would be observed with $(B-V) < 0.01$, whereas 16 are
seen. In addition, any stars evolving from the most-populous portion
are predicted to evolve brighter than the observed red end of the HB
by 0.2 mag or more. The difficulties are even greater for stars
evolving from the blue end of the HB - they spend a smaller fraction
of their time near the horizontal part of the HB, and evolve to the
red (if at all) at even higher luminosity. A decreased helium
abundance would constrain the HB evolutionary tracks to be closer to
the zero-age HB, but would also tend to shift the HB stars as a group
to the red. The helium indicators $R$ and $\Delta$ do not indicate
that NGC 288's helium is any different from the other clusters in this
group.

If we assume the difference to be due to age alone, then we find that
NGC 288 is approximately $3.7 \pm 1.5$ Gyr older than M5 (using
$\partial M_{V}(\mbox{TO}) / \partial t_{age} \approx 0.07 \
\mbox{mag} / \mbox{Gyr}$ from BV92 isochrones). We conclude that, to
within the errors, six of the clusters in this metallicity group (NGC
362, NGC 1261, NGC 1851, NGC 2808, M4, and M5) appear coeval to an
accuracy of about $\pm1.5$ Gyr. Two other clusters may have different
ages: NGC 288, which may be older, and Palomar 5, which may be
younger. If we leave these two clusters out of consideration, we are
still left with a large variation in HB morphology between M5 and
either NGC 362 or 1261. Because the sensitivity of HB morphology to
age depends on the unknown mass loss mechanism (Lee, Demarque \& Zinn
1994), the viability of age as the second parameter varies. For a
constant mass difference between helium flash red giants and zero-age
HB stars, age differences of approximately 4 to 5 Gyr are required to
explain the morphological variations --- a case that can be ruled out
by this data. If a Reimers' mass loss relation holds, age differences
of only about 2 to 3 Gyr are required. This possibility cannot be
ruled out.

\subsubsection{Isochrone and LF Comparison}\label{isolf}

The most direct absolute age determination that can be made involves
comparing the absolute magnitude of the turnoff to models. Using the
subdwarf distance modulus, we find $M_{V}(TO) = 4.16 \pm 0.08$, where
the error refers only to the formal errors in the distance modulus and
in the determination of the apparent magnitude of the turnoff.
Figure~\ref{fig29} places this point in a comparison with turnoffs
from the BV92 isochrones (which implicitly assume bolometric
corrections from VandenBerg 1992). It should be noted that
$\alpha$-element enhancements shift the theoretical relations to lower
metallicity, and hence result in a slight ($\sim$ 1 Gyr) age reduction
for clusters on the whole. With $\alpha$-enhancement included, we find
an empirical relationship
\[ \log_{10} t_{9} = -0.874 - 0.118 \ \mbox{[$M$/H]} + 0.446
M_{V}(\mbox{TO}) ,\] which becomes
\[ \log_{10} t_{9} \approx -0.874 - 0.118 \ \mbox{[Fe/H]} - 0.08 \
[\alpha/\mbox{Fe}] + 0.446 M_{V}(\mbox{TO}) . \] Via this relation, we
find an age of $14 \pm 1.2$ Gyr.  Determinations of $M_{V}$(TO) are
also plotted for the clusters M3 (Buonanno et al. 1994), M15 (Durrell
\& Harris 1993), and M92 (Stetson \& Harris 1988, as redetermined by
Bolte \& Hogan 1995). If the zero-point of the observed metallicity
scale is actually that favored by the Lick-Texas group, then for M5,
we derive $M_{V}(TO) = 4.07\pm0.09$, and an age of $11.6\pm1.1$ Gyr.
On the ZW84 metal scale, the ages of these clusters are consistent
with each other to within the errors. However, the Lick-Texas scale
would imply age differences: M92 has almost identical values in both
studies, M15 is 0.15 dex lower, and M3 and M13 are about 0.2 dex
higher on the Lick-Texas scale.

Finally, we compare the M5 fiducial with the isochrones of BV92.
Although the model difficulties with predicting colors are well known
(e.g. VandenBerg, Bolte \& Stetson 1996), there are still valuable
cross-checks to be made in comparing the observations to model
isochrones throughout the CMD. Figure~\ref{fig30} shows 14 Gyr
isochrones for a range of metal contents, with the M5 fiducial {\it
overlaid} after being shifted in color to correct for the reddening
(E$(B-V) = 0.03\pm0.01$), and in magnitude to remove the apparent distance
modulus ($(m - M)_{V} = 14.41\pm0.07$ as derived in \S
\ref{subd}). Not surprisingly the models and data agree very well
along the unevolved main sequence --- after all, the subset of
subdwarfs that we used to calibrate the distance modulus had
photometric values very close to the theoretical predictions. However,
the shapes of the isochrones also match quite well for the 14 Gyr
isochrones, up to the upper RGB, which is blue relative to the
isochrones. The $\alpha$-enhanced isochrones shown in
Figure~\ref{fig32} match the M5 fiducial better overall. (This seems
to be evidence that $\alpha$-enhancements do have observable effects
on the shape of the cluster fiducial line.) From the comparison to the
$\alpha$-enhanced isochrones, we infer an age of between 12 and 14 Gyr.
There is reason to hope that the external errors associated with this
kind of fit have decreased in the past few years. As we saw in \S
\ref{met}, there has been little change in the shape of the isochrones
since the last set of oxygen-enhanced isochrones was released, and
there have only been small shifts absolutely in color and magnitude.

As Renzini \& Fusi Pecci (1988) pointed out, examining the union of
CMD and LF data is the most effective way of making detailed
comparisons with theoretical models --- testing our ideas about
stellar structure and timescales simultaneously. In considering age
though, a LF comparison can be more sensitive than an isochrone
comparison if the proper filter band is chosen. $B$-band is ideal for
intermediate metal-poor clusters ($-1.4 \lesssim$ [Fe/H] $\lesssim
-0.9$), as the SGB has nearly constant magnitude in this
range. Consequently, small changes in age (or chemical composition)
result in slight changes in the slope of the SGB, but significant
changes in the number of stars falling in the ``SGB peak'' feature in
the LF. With a well-determined composition and distance modulus, this
feature can provide excellent age discrimination. (In a similar vein,
$V$-band LFs provide the best age sensitivity for more metal-rich
clusters.)

Figure~\ref{fig31} shows the comparison of the $B$-band LF with the
theoretical LFs in the vicinity of the SGB using the apparent distance modulus
$(m - M)_{V} = 14.42$ for [$M$/H] $= -1.26$. The [$M$/H] $= -1.26$
model roughly matches the position of the subgiant branch ``jump'',
although it does not reproduce the star counts in the same range for
an age of 14 Gyr. A small decrease in helium abundance would improve
agreement between the theoretical and observed LFs, as helium changes
affect only the SGB portion of the LF (see \S
\ref{sgblf}). Alternately, an age of approximately 13 Gyr would bring
the models into better agreement with the observed LF in both
magnitude position and size.

We must also consider the possibility of alternate metal contents for
M5.  For models having [$M$/H] $= -1.48$, it is not possible to
produce an SGB peak of the observed size, even for ages as low as 10
Gyr. However, we have to consider seriously models having a metal content of
[$M$/H] $= -1.03$ ([Fe/H] $=1.17$ and $\alpha$-enhancements) for M5 as
this is the value measured based on modern, high-dispersion
spectroscopy.  Models having a metal content of [$M$/H] $= -1.03$ are able to
approximately reproduce the peak height with ages between about 9 and
15 Gyr.  For this metal content and an age of 14 Gyr, the position of
the SGB feature in magnitude would imply a distance modulus 0.3 mag
smaller than derived from subdwarf fitting, as can be seen in
Figure~\ref{fig13}. An age of approximately 10 Gyr would be required
to match the magnitude position of the SGB, and the height of the
peak. This, however, is inconsistent with the color extent of the SGB
in the theoretical isochrones. (Given the sensitivity of the SGB extent
to the choice of mixing-length parameter, this objection to the
combination of low age and high metallicity could be weakened if
the mixing-length theory is inadequate.) Independent of the distance modulus
used, the size of the SGB peak strongly rules out ages of greater than
15 Gyr, and metal contents [$M$/H] $\lesssim -1.3$ ([Fe/H] $\lesssim
-1.5$).

With regards to the age determination, the uncertainties in the the
metallicity scale in the range around [Fe/H]=--1.3 is an important
issue. For the Zinn \& West scale, with [Fe/H] $=-1.40$, there is
nearly complete consistency between the M5 CMD and LF observations and
the BV92 models for an age between 13 and 14 Gyr with the LF providing
the lower age. (An important exception is the different distance
moduli derived from subdwarf and HB fitting.)  Whether this is
significantly different than the age of the well-studied more
metal-poor M92 is difficult to answer. Bolte \& Hogan (1995) use
essentially the same subdwarf list and parameters to derive a distance
to M92 and and age based on BV92 of 15.8 Gyr. However, if the
metallicity scale of the Lick-Texas group, which is the same as Zinn
\& West for M92 but 0.23 dex more metal-rich for M5, is used, then the
derived age for M5 is reduced to 11.5 Gyr. For the higher metallicity,
low-age models, the consistency between the models and observations is
not as complete, however. The inconsistencies are greatest in the
areas where the models have weaknesses -- prediction of the luminosity
of the RGB bump, and the color extent of the SGB.  The {\it relative}
metallicities of M5 and other clusters of similar abundance are well
established, as is the similarity in the ages of the clusters
considered in \S \ref{delv}.  If the higher [Fe/H] values turn out to
be the correct ones, a strong case can be made for an age-metallicity
relationship in the halo globulars.

\section{Conclusions}\label{7.}

1. The population ratios derived from the bright star sample
(complete into the center of the cluster) indicate
that M5 stars have a semiconvective zone during their HB evolution.
More puzzling though, is the indication that the $R$ and $R^{\prime}$ ratios
reflect a helium abundance of $Y = 0.19 \pm 0.02$. This value is
is seen at all cluster radii.

We recalibrate the helium indicator $\Delta$ and reexamine $A$ using
recent theoretical models. Overall, these indicators appear to
show that the clusters in
M5's metallicity group (including M5) have helium abundances
consistent with each other and with $Y \approx 0.23$. By the indicator
$R$, the clusters are consistent with each other to within the errors,
but appear to have values indicating low helium abundances.
More globally, $\Delta$ shows no apparent variation with [Fe/H], while
$R$ and $A$ appear to increase with decreasing [Fe/H] for the limited sample
of clusters examined. We recommend that $\Delta$ be
used as the helium indicator of choice for Galactic globular clusters
with HB morphologies populating the instability strip,
so as to overcome the statistical limitations
inherent in $R$ and the time-series photometry necessary to compute $A$.

2. Extensive artificial star tests have been conducted to determine
the incompleteness of our sample of stellar photometry as functions of
magnitude and radius.
Blending effects have been avoided by restricting the LF sample to
stars well outside the core of the cluster.
When blends are corrected for in the $B$-band LF, peak and
dip features become apparent on the SGB. The $I$-band LF allows us to
examine subgiant evolutionary time-scales in detail, and no
significant deviations from theoretical predictions are found.

3. The relative star-count levels of the LF on the MS and RGB agree
with theoretical predictions, in contrast to the case for metal-poor
clusters like M30 and M92. In $I$-band, there may be a slope
difference between the observations and theoretical predictions for
the RGB. There is slight evidence for a deficit of the brightest giants in the
core of the cluster relative to the outskirts.

4. The RG bump is observed to be fainter than the theoretical LFs
indicate, in agreement with observations of the bump in several other
clusters. However, once the effect of $\alpha$-element enhancement is
taken into account, this discrepancy is significantly reduced. This
decreases the need for non-standard mixing mechanisms (such as
convective overshooting or meridional circulation) during first
dredge-up on the RGB to force a match between theory and observations.

5. We identify 8 new RR Lyrae candidates in the core of M5 based on a
series of $V$-band exposures, and the apparent scatter in their
measured values on these frames.

6. We confirm that the metallicity of M5 is approximately the same as
NGC 288 and NGC 362 (and other clusters compared listed in ZW84 as
having [Fe/H] $\approx -1.35$), based on photometric
indicators. However, we are unable to make a firm distinction between
the competing values of the {\it absolute} metallicity: [Fe/H] $= -1.40$ (ZW84)
or $-1.17$ (Sneden et al. 1992). With recent indications that the ZW84
scale may be nonlinear, this question should be pursued further.
However, an enhancement in the $\alpha$-element abundances for the
cluster definitely brings spectroscopic measurements of the metallicity into
better agreement with indications from isochrone and LF fitting. 

Isochrones and LFs for $\alpha$-enhanced compositions appear to fit
the observed fiducial line and LFs better than oxygen-enhanced
models. In addition, the colors of the ``classical subdwarfs'' with
trigonometric parallaxes are predicted accurately when
$\alpha$-element enhancements are taken into account.

7. The apparent distance modulus for M5 derived from subdwarf fitting of the MS
is $(m - M)_{V} = 14.41\pm0.07$ for [Fe/H] $= -1.40$ ($(m - M)_{V} =
14.50\pm0.07$ for [Fe/H] $= -1.17$). We include a list of 23 subdwarfs
(and 5 subgiants) that are likely to be useful in future
determinations of the distance modulus, once their parallaxes and
metallicities are better determined. The excellent agreement of the
best measured subdwarf, Groombridge 1830 (HD 103095), with theoretical
isochrones indicates that the distance modulus should be quite well
determined aside from uncertainty in the metallicity. The derived
distance modulus also agrees well with the value from a
Baade-Wesselink analysis of RR Lyraes by Storm et al. (1994).

A model fit to the horizontal branch gives an apparent distance modulus $(m -
M)_{V} = 14.52\pm0.05$. This appears to indicate a systematic
difference between values taken from subdwarf fitting and from some HB
model fits. The AGB (especially the AGB clump) is not well fit by the
models of Dorman (1992b), but appears to be consistent with those of
Castellani et al. (1991).

8. From isochrone and LF fitting considerations, M5 has an age of
approximately 13.5 Gyr with an internal error of $\pm 1$ Gyr. Our age
estimate would decrease by approximately 2 Gyr if the Lick-Texas
zero-point of the metallicity scale is taken instead of the ZW84 value. M5
also appears to have approximately the same age as most of the other
clusters of comparable metallicity, according to calculations of $\Delta
V_{TO}^{HB}$. This constraint appears to be capable of eliminating age
as the second parameter in HB morphology in the case of constant mass
loss between RGB and HB (although a Reimers' mass-loss relation may
still be valid). NGC 288 is a possible exception, as it
appears to be approximately $3.7 \pm 1.5$ Gyr older. Considerable
differences between NGC 288's metallicity or helium abundance
and those of the other clusters considered cannot explain the
differences in $\Delta V_{TO}^{HB}$. Evolution of the reddest HB stars
away from the zero-age HB is ruled out by comparison with theoretical
evolutionary tracks and timescales.

\acknowledgments

We would also like to thank J. Faulkner, R. Hanson, R. Kraft, R. Rood, 
and M. Shetrone for very helpful discussions. We would especially like to
thank D. VandenBerg for the use of preliminary $\alpha$-enhanced
isochrones, and the referee (B. Carney) for a careful reading and for
suggestions that improved the paper a great deal.
This research has made use of the SIMBAD database, operated at CDS,
Strasbourg, France. One of us (ELS) was partially supported under a
National Science Foundation Graduate Research Fellowship.

Electronic copies of the listings of the photometry for the $BI$ and
$BVI$ samples are available on request to the first author.

\newpage

\figcaption{(a) Final magnitude residuals as a function of magnitude for
the comparison of Landolt standards and
measured values in this study. The residuals are in the sense of (us
- Landolt). Stars plotted with $\times$ were not
included in the transformation equation calibrations. (b) Magnitude
residuals as a function of color. (c) Color residuals as a function of
color. \label{fig1}}

\figcaption{ $V$-band finding chart for calibrated secondary standards in M5.
\label{fig2}}

\figcaption{ Final residuals for the comparison of our M5 secondary
standards with those of Stetson (1994b). \label{fig3}}

\figcaption{ Residuals for the comparison between the CCD photometry of
Storm, Carney \& Beck (1991) and the CTIO 4m data. \label{fig4}}

\figcaption{ Residuals for the comparison between the photoelectric
photometry of Lloyd Evans (1983) and the CTIO 4m data. \label{fig5}}

\figcaption{ A comparison of fiducial lines for several studies of M5.
Crosses ($\times$) and circles ($\bigcirc$) are derived from the inner
and outer field photometry of Richer \& Fahlman (1987), respectively.
Triangles ($\bigtriangleup$) come from Buonanno et al. (1981), and
squares ($\Box$) from Simoda \& Tanikawa (1970). The solid line is
fiducial derived from the current dataset, and is unsmoothed. \label{fig6}}

\figcaption{ The color-magnitude diagram for all objects in the total
(a) $BVI$ sample and (b) $BI$ sample. \label{fig7}}

\figcaption{ The color-magnitude diagram for samples restricted to
stars having low CHI values. (a) The $BVI$ sample. (b) The $BI$
sample. \label{fig8}}

\figcaption{ Internal magnitude errors versus instrumental $B$ magnitude.
Circles indicate median calculations for magnitude bins, and the solid
line shows the calculated fitting function. The dotted lines indicate
the boundary curves used to estimate the uncertainty in the magnitude
error. \label{fig9}}

\figcaption{ Magnitude bias versus instrumental $B$ magnitude.
\label{fig10}}

\figcaption{ Total detection probability versus instrumental $B$
magnitude. \label{fig11}}

\figcaption{ External magnitude errors versus instrumental $B$
magnitude. \label{fig12}}

\figcaption{ The $B$-band luminosity function in the case of extreme
crowding, compared to our best luminosity function ({\it dashed
line}). The peak feature on the SGB has been erased as a result of
blending, and an additional bump has formed on the lower RGB. A
theoretical luminosity function of lower metallicity ({\it solid
line}) is a better fit to this data. \label{fig15}}

\figcaption{ The bright sample of M5, including RR
Lyrae stars with intensity-weighted $V$ magnitudes and
magnitude-weighted $(B-V)$ colors from Storm, Carney \& Beck (1991).
The sample has been restricted to stars with
projected radii greater than $100^{\prime\prime}$ from the cluster
center. \label{fig21}}

\figcaption{ A comparison of red giant branches in the $(M_{I},V-I)$
plane. From left to right at $M_{I} = -3.4$, the clusters picture are:
M15, NGC 6397, M2, NGC 6752, NGC 1851, M5 ({\it dotted line}), and 47
Tucanae. \label{fig22}}

\figcaption{ A comparison of oxygen-enhanced isochrones ({\it solid
lines}; Bergbusch \& VandenBerg 1992) with a preliminary
$\alpha$-enhanced isochrone with [Fe/H] = --1.31 and
[$\alpha$/Fe] = +0.3 ({\it dotted line}; VandenBerg 1995) for an age of
14 Gyr. From right to left, the metallicities of the
oxygen-enhanced isochrones are: [Fe/H] = --0.78, --1.03, --1.26, --1.48,
--1.66, --1.78, --2.03, and --2.26. \label{fig23}}

\figcaption{ A comparison of luminosity functions for oxygen-enhanced
compositions ({\it solid lines}; Bergbusch \& VandenBerg 1992)  with
a preliminary luminosity function for an $\alpha$-enhanced composition
of [Fe/H] = -1.31 and
[$\alpha$/Fe] = +0.3 ({\it dotted line}; VandenBerg 1995) for an age of 14 Gyr.
At the vertical subgiant branch feature from left to right, the
metallicities of the oxygen-enhanced luminosity functions are: [Fe/H]
= --1.48, --1.26, and --1.03. The luminosity functions all have the same
mass function slope $x = -0.5$. \label{fig24}}

\figcaption{ A plot of all stars included in the bright-star population
study for which we have calibrated magnitudes from the CTIO 4m data.
Circles are HB stars, squares are AGB stars, and triangles are RGB
stars. \label{fig17}}

\figcaption{ The color-magnitude diagram for the uncalibrated CFHT
data. \label{fig18}}

\figcaption{ The cumulative radial distribution for AGB stars ({\it
solid line}), HB stars ({\it dotted line}), and RGB stars ({\it dashed
line}). The sample used extends
from the center of the cluster to the edge of the CTIO 4 m frames.
\label{fig19}}

\figcaption{ Published fiducial lines for the horizontal branches of
several clusters: M5 ({\it solid line}; this paper), NGC 288 ({\it
dotted line}; Bolte 1992), NGC 362 ({\it short dashed line}; Harris
1982), NGC 1851 ({\it long dashed line}; Walker 1992a), and M3 ({\it
dot-dashed line}; Buonanno et al. 1994). The fiducial lines for each
cluster have been shifted by amounts equal to those used in the
relative age method of VandenBerg, Bolte \& Stetson (1990). \label{fig20}}

\figcaption{ The subdwarf sample after correction to a common
metallicity [Fe/H] $= -1.19$.  The absolute magnitudes have been adjusted for
Lutz-Kelker bias. ``Classical'' subdwarfs ($\bullet$) and
the extended list ($\Box$) are plotted. Bergbusch \& VandenBerg (1992)
oxygen-enhanced isochrones ({\it solid lines}) for 14 Gyr are plotted for
comparison, with the same metallicity spread as in Figure \ref{fig23}. The fit
to the fiducial line of M5 ({\it dotted line}) is shown. Only the
classical subdwarfs were used in the fit. \label{fig25}}

\figcaption{ Color difference between photometry of subdwarf
(corrected for $\alpha$-element enhancement) and the prediction from
the Bergbusch \& VandenBerg (1992) isochrones, as a function of
metallicity. The symbols are in the same sense as Figure \ref{fig25}.
\label{fig26}}

\figcaption{ The $B$-band luminosity function for M5, with
theoretical luminosity functions having ages of 14 Gyr. The
theoretical LFs have been adjusted in distance modulus for best fit to
the subgiant jump. \label{fig27}}

\figcaption{ The fit to the observed horizontal branch stars having
projected radii greater than $100^{\prime\prime}$ from the cluster
center. The theoretical zero-age horizontal branches and horizontal
branch tracks are from Dorman
(1992b). From left to right, the panels show the models for [Fe/H] =
--1.03, --1.26, and --1.48. The models have helium abundance
$Y_{HB} \approx 0.25$, corresponding to $Y_{MS} \approx
0.236$. \label{fig28}}

\figcaption{ The $B$-band luminosity function for M5, with
theoretical luminosity functions having ages of 14 Gyr. The
theoretical LFs have been adjusted in distance modulus to agree with
values derived from subdwarf fitting. \label{fig13}}

\figcaption{ The $I$-band luminosity function for M5. Theoretical
models have been adjusted as in Figure \ref{fig13}. \label{fig14}}

\figcaption{ The $B$-band luminosity function on the red giant branch,
centered on the RGB bump. The distance moduli are derived from
subdwarf fitting. The turnups at the bright ends of the theoretical
LFs are caused by bolometric corrections, which force the giant branch to
become horizontal. For $B$-band, this is near the tip of the RGB.
\label{fig16}}

\figcaption{ The age of M5 as derived from the position of the MS
turnoff, along with the uncertainties. The solid line represents a fit
to turnoff data from the isochrones of Bergbusch \& VandenBerg (1992),
with an additional correction for an $\alpha$-element enhancement of
+0.3 dex. The fit represented by the dotted line does not include this
correction. The open circle represents the value derived for the
higher Lick-Texas value of [Fe/H] for M5. \label{fig29}}

\figcaption{ A comparison of the M5 fiducial ({\it dotted line}) with
oxygen-enhanced isochrones ({\it solid lines}) in the $(M_{V},B-V)$
plane (VandenBerg 1995). From left to right, the 14 Gyr isochrones
have [Fe/H] = --1.48, --1.26, and --1.03. \label{fig30}}

\figcaption{ A comparison of the M5 fiducial ({\it dotted line}) with
$\alpha$-enhanced isochrones ({\it solid lines}) in the $(M_{V},B-V)$
plane (VandenBerg 1995). From left to right, the [Fe/H] = --1.31
([$\alpha$/Fe] $= +0.3$) isochrones
have ages 12, 14, and 16 Gyr. The apparent distance modulus used here is $(m -
M)_{V} = 14.50$. \label{fig32}}

\figcaption{ A comparison of the M5 $B$-band LF with
oxygen-enhanced models (VandenBerg 1995). Each of the theoretical
models has [Fe/H] $= -1.26$ and apparent distance modulus $(m-M)_{B} =
14.45$. \label{fig31}}

\end{document}